\documentclass[reprint,aps,longbibliography,nofootinbib,floatfix,nobalancelastpage]{revtex4-2}
\usepackage{amsmath}
\usepackage{amsfonts}
\usepackage{amssymb}
\usepackage{mathtools}
\usepackage{graphicx}
\usepackage[dvipsnames]{xcolor}
\usepackage[colorlinks=True,citecolor=gray,linkcolor=myRed,urlcolor=gray]{hyperref}
\usepackage{hypcap}
\usepackage{yfonts}
\usepackage{bm}
\usepackage{ulem}
\usepackage{dsfont}
\usepackage{physics}
\newcommand\eq[1]{\begin{align}#1\end{align}}

\newcommand\nh{N_\mathcal{H}}

\newcommand{\pu}{\mathcal{P}^{AB}}
\newcommand{\puo}{\tilde{\mathcal{P}}^{AB}}
\newcommand{\pudyn}{\tilde{\mathcal{P}}_{{\rm dyn}}}
\newcommand{\pusat}{\tilde{\mathcal{P}}_{{\rm sat}}}
\newcommand{\pusq}{\tilde{\mathcal{P}}_\square}
\newcommand{\purarb}{\tilde{\mathcal{P}}_{\square_{r_Ar_B}}}
\newcommand{\pur}{\tilde{\mathcal{P}}_{\square_r}}

\newcommand{\al}{\alpha}
\newcommand{\be}{\beta}
\newcommand{\ga}{\gamma}
\newcommand{\la}{\lambda}
\newcommand{\ta}{\theta}

\newcommand{\Vabcd}{V_{\alpha\beta\gamma\lambda}}

\newcommand{\Vabab}{V_{\alpha\beta\alpha\beta}}

\newcommand{\tabcd}{\theta_{\alpha\beta\gamma\lambda}}

\definecolor{myBlue}{RGB}{31,119,180}
\definecolor{myOrange}{RGB}{255,127,14}
\definecolor{myGreen}{RGB}{44,160,44}
\definecolor{myRed}{RGB}{214,39,40}
\definecolor{myPurple}{RGB}{148,103,189}

\makeatletter
\def\p@figure{\color{myBlue}}
\def\p@equation{\color{myRed}}
\makeatother

\begin{document}

\title{Entanglement dynamics and eigenstate correlations in \\strongly disordered quantum many-body systems}

\author{Bikram Pain}
\email{bikram.pain@icts.res.in}
\affiliation{International Centre for Theoretical Sciences, Tata Institute of Fundamental Research, Bengaluru 560089, India}

\author{Sthitadhi Roy}
\email{sthitadhi.roy@icts.res.in}
\affiliation{International Centre for Theoretical Sciences, Tata Institute of Fundamental Research, Bengaluru 560089, India}

\begin{abstract}
The many-body localised phase of quantum systems is an unusual dynamical phase wherein the system fails to thermalise and yet, entanglement grows unboundedly albeit very slowly in time. We present a microscopic theory of this ultraslow growth of entanglement in terms of dynamical eigenstate correlations  of strongly disordered, interacting quantum systems in the many-body localised regime. These correlations involve sets of four or more eigenstates and hence, go beyond correlations involving pairs of eigenstates which are usually studied in the context of eigenstate thermalisation or lack thereof. We consider the minimal case, namely the second R\'enyi entropy of entanglement, of an initial product state as well as that of the time-evolution operator, wherein the correlations involve quartets of four eigenstates. We identify that the dynamics of the entanglement entropy is dominated by the spectral correlations within certain special quartets of eigenstates. We uncover the spatial structure of these special quartets and the ensuing statistics of the spectral correlations amongst the eigenstates therein, which reveals a hierarchy of timescales or equivalently, energyscales. We show that the hierarchy of these timescales along with their non-trivial distributions conspire to produce the logarithmic in time growth of entanglement, characteristic of the many-body localised regime. The underlying spatial structures in the set of special quartets also provides a microscopic understanding of the spacetime picture of the entanglement growth. The theory therefore provides a much richer perspective on entanglement growth in strongly disordered systems compared to the commonly employed phenomenological approach based on the $\ell$-bit picture.
\end{abstract}

\maketitle

\section{Introduction}

The question of if, and how, out-of-equilibrium quantum many-body systems thermalise as quantum correlations develop dynamically between distant degrees of freedom is one of the most fundamental questions of interest in condensed matter and statistical physics~\cite{deutsch1991quantum,srednicki1994chaos,tasaki1998from,rigol2008thermalisation,dalessio2016from,nandkishore2015many}.
This question is intimately connected to the entanglement structure and its dynamics~\cite{popescu2006entanglement,linden2009quantum,kaufman2016quantum} as it is one of the most fundamental manifestations of how quantum correlations and information propagates through the system~\cite{clalabrese2005evolution,bardarson2012unbounded,serbyn2013universal,kim2013ballistic,nanduri2014entanglement,lakshminarayan2016entanglement,nahum2017quantum,nahum2018dynamics,huang2021extensive}.
Developing a microscopic theory for the dynamics of entanglement in quantum many-body systems is therefore naturally important.

Given a Hamiltonian (or the generator of time translation in general), the joint distribution of its eigenvalues and eigenstates, in principle, contains all the information about the dynamics of the system. However, with the Hilbert-space dimension growing exponentially with the system size, it is obvious that the amount of information in the joint distribution is too large for it to be feasible to develop a theory in terms of that.
A key theoretical challenge then is to distil out the minimal correlations between the eigenvalues and eigenstates that suffice to develop a theory for the dynamics of local observables and entanglement. 
As far as the former is concerned, the eigenstate thermalisation hypothesis (ETH)~\cite{dalessio2016from,deutsch2018eigenstate} provides a statistical description of the matrix elements of local operators in the eigenbasis and sheds light on their dynamics. An important understanding from the ETH is that local observables do eventually thermalise in systems which satisfy the ETH~\cite{rigol2008thermalisation,beugeling2014finitesize,alba2015eigenstate,luitz2016anomalous,borgonovi2016quantum,dalessio2016from,deutsch2018eigenstate,roy2018anomalous,grover2018does,khaymovich2019eigenstate}. On the other hand, systems which violate the ETH, such as strongly disordered systems in the many-body localised regime (MBL)~\cite{oganesyan2007localisation,znidaric2008many,pal2010many,nandkishore2015many,abanin2017recent,abanin2019colloquium,alet2018many,sierant2024manybody} fail to thermalise and break ergodicity.

While the ETH or its violation is central to our understanding of dynamics of local observables, it has been recognised that the picture is insufficient. It has been established that there exist non-trivial higher-point correlations between the eigenvalues and eigenstates which fall outside the purview of the ETH, and more importantly, which encode the dynamics of information scrambling and entanglement growth~\cite{foini2019eigenstate,chan2019eigenstate,brenes2021out,pappalardi2022eigenstate,shi2023local,hahn2023statistical,jafferis2023matrix,pappalardi2023general,jindal2024generalized}.
In fact, for locally interacting systems, the question can be turned around and posed as, what does the presence of fundamental bounds on how fast quantum information can propagate~\cite{lieb1972finite} imply for higher-point, dynamical eigenstate correlations. However, much of the literature on this is focused on ergodic or quantum chaotic systems and hitherto, there is no work addressing such questions for non-ergodic systems. 
This is the central motivation of this work, namely,
to understand the entanglement dynamics in strongly disordered quantum systems, in the  MBL regime, through the lens of eigenstate correlations.

Within the rich dynamical phase diagram of disordered, interacting quantum systems, the MBL phase constitutes a rather interesting but unusual example of a robustly non-ergodic phase which violates the ETH~\cite{oganesyan2007localisation,znidaric2008many,pal2010many,nandkishore2015many,abanin2017recent,abanin2019colloquium,alet2018many,sierant2024manybody} and yet shows unbounded growth of entanglement~\cite{znidaric2008many,bardarson2012unbounded,serbyn2013universal,nanduri2014entanglement,chen2016universal,chen2017out,huang2017out,deng2017logarithmic}, albeit logarithmically slowly in time.
This behaviour sets the MBL phase apart not only from the ergodic phase, but also from a non-interacting Anderson localised phase~\cite{anderson1958absence} wherein there is no entanglement growth. 
An early and simplistic understanding of this logarithmic growth of entanglement in the MBL phase was via the phenomenological, so-called $\ell$-bit picture~\cite{serbyn2013local,huse2014phenomenology}. 
The picture proposes that there exists an extensive number of (quasi)local integrals of motion, the $\ell$-bits, which are weakly dressed versions of the trivially localised integrals of motion at infinite disorder.
Interactions between these $\ell$-bits decay exponentially in space such that degrees of freedom separated by a certain distance get entangled on timescales which are exponentially large in the distance, which eventually leads to the logarithmic growth of entanglement~\cite{bardarson2012unbounded,serbyn2013local,serbyn2013universal}.

While the $\ell$-bit picture has been lead to several important insights about the MBL phase, their explicit constructions have remained elusive despite several noteworthy efforts~\cite{ros2015integrals,chandran2015constructing,pekker2017fixed}. As a result, a microscopic understanding of the distribution of the $\ell$-bits' localisation lengths, which would be an important ingredient to any theory, has also eluded us. More recently, it has been realised that many-body resonances between the $\ell$-bit configurations abound the spectra of MBL systems~\cite{morningstar2021avalanches,garratt2021resonances,sels2021dynamical,garratt2022resonant,crowley2022constructive,sels2023thermalisation} and hence the phenomenological $\ell$-bit picture cannot be entirely complete. These difficulties with the $\ell$-bit picture therefore underline the importance of a microscopic theory of the logarithmically slow entanglement growth in the MBL phases, without alluding to any phenomenological picture.

In this work, we approach this question from the point of view of eigenstate correlations. 
The precise questions raised therefore are what are the minimal eigenstate correlations that encode the dynamics of entanglement, what are their statistical properties in MBL systems, and how do they lead to the ultraslow growth of entanglement with time. A detailed understanding of these questions constitutes the main result of this work.

\tableofcontents

\subsection{Overview of the main results}
We start with a brief overview of the main results of the work. As a measure of entanglement in the system, we consider the ($q^{\rm th}$-)R\'enyi entropy of entanglement between a subsystem $A$ and its complement $B$, defined as
\eq{
S^{AB}_q(t) = -\frac{1}{q-1}\ln{\rm Tr}[\rho_A^q(t)]\,,
}
where $\rho_A(t)$ is the reduced density matrix of $A$ at time $t$. In particular, we consider the simplest case of the second R\'enyi entropy, $S^{AB}_{q=2}(t)$, or equivalently $\pu(t) = {\rm Tr}\rho_A^2(t)$\,, where $\pu(t)$ is defined as the purity of $\rho_A(t)$. Throughout the work we consider one-dimensional systems where $A (B)$ is the left (right) half of the system.

We identify that averaging over random product states between $A$ and $B$ as initial states allows for the time-dependent subsystem purity to be expressed in terms of sums of eigenstate and spectral correlations involving four eigenstates. In fact, with this averaging the time-dependent subsystem purity is identical to the operator purity of the time-evolution operator. As such, our theory describes on equal footing the dynamics of entanglement of states starting from typical product states and the operator entanglement of the time-evolution operator.
These correlations between quartets of eigenstates form the building blocks of our theory.
We show that resolving these correlations in frequency ($\omega$) where the frequency is now a combination of the four eigenvalues, directly encodes the dynamics of the subsystem purity.
In particular, we find that the frequency dependence of these correlations is a power-law which in turn implies a power-law decay in time of $\pu(t)$ and consequently, the logarithmic growth in time of $S^{AB}_2(t)$. 
Note that since these correlations involve four eigenstates, they manifestly go beyond the question of the ETH or its violation. Although in our case, these correlations emerge naturally out the dynamics of subsystem purity, their closely related cousins have been studied extensively for ergodic systems in the context of operator entanglement entropy of the time-evolution operator~\cite{hahn2023statistical}.

Having distilled out the minimal correlations that encode the dynamics of entanglement we next turn to the question of the microscopic origins of the power law in time of $\pu(t)$ or equivalently, the power law in $\omega$ of its Fourier transform, $\puo(\omega)$. Insights into the question are obtained by studying the anatomy of the four-point eigenstate correlations in detail. For a system with total Hilbert-space dimension $\nh$, there are obviously $O(\nh^4)$ possible quartets of eigenstates. 
Quite remarkably, we find that only $O(\nh^2)$ of those quartets carry an $O(1)$ value of the relevant eigenstate correlation whereas for the rest of them, the correlations are vanishingly small.
The key point is here that, for the latter, the correlations are so small that even their overwhelmingly large majority does not lead to a contribution comparable to the former.
As such, we conclude that the dynamics of purity is dominated entirely by the $O(\nh^2)$ set of `special dominant' quartets of eigenstates for which the eigenstate correlation is $O(1)$. In fact, we find that once restricted to these special quartets, the {\it spectral correlations} within them are sufficient to recover the aforementioned power-laws in time and $\omega$.
The question then reduces to understanding the structure of these special quartets and the spectral correlations within them.

\begin{figure}[!t]
\includegraphics[width=\linewidth]{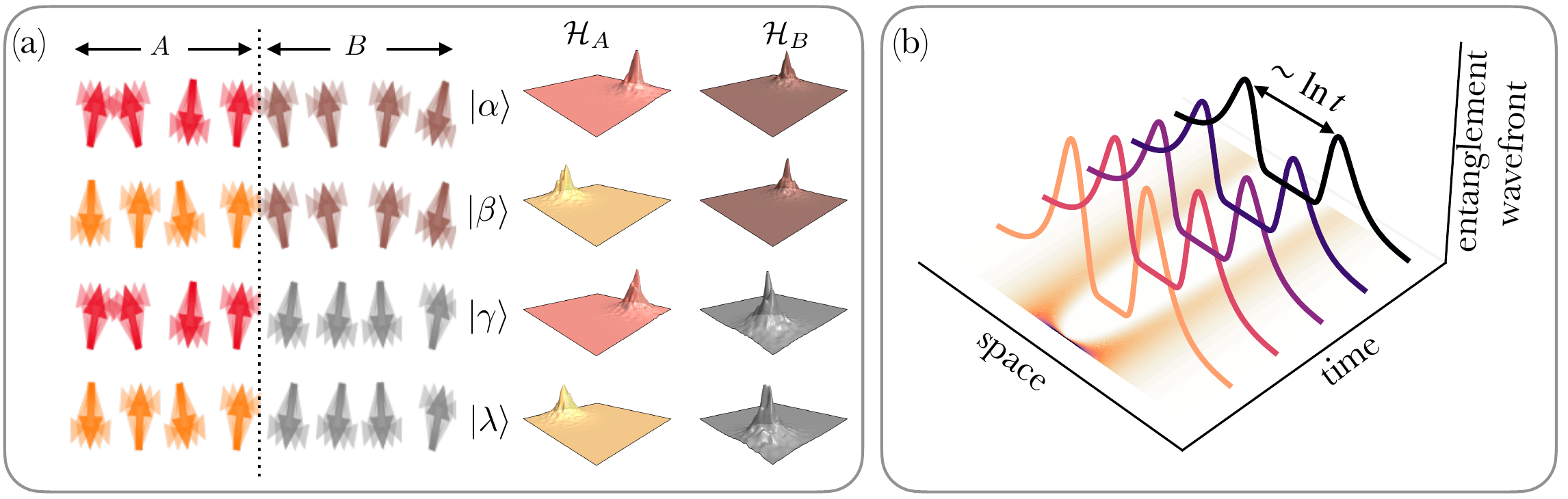}
    \caption{Schematic figure summarising the central results of the work. (a) A cartoon of the special dominant quartets, eigenstate and spectral correlations between which dominate the entanglement growth in a many-body localised system. The four eigenstates are denoted as $\{\ket{\alpha},\ket{\beta},\ket{\gamma},\ket{\lambda}\}$. The defining feature of these quartets is that for any eigenstate in the quartet, say $\ket{\alpha}$ there exists two others, say $\ket{\beta}$ and $\ket{\gamma}$ which differ from $\ket{\alpha}$ only in subsystem $A$ and $B$ respectively whereas the fourth one $\ket{\lambda}$ differs from $\ket{\alpha}$ in both subsystems. The left column shows the cartoon in real space for a spin chain with the spins approximately polarised. The right column shows schematically the wavefunctions corresponding to the eigenstates in the Hilbert subspaces of subsystems $A$ and $B$. (b) Schematic figure showing the logarithmically spreading entanglement wavefront in spacetime. The wavefront indicates that at any time $t$, the entanglement growth is dominated by the dynamics at a distance $r\sim \ln t$ from the bipartition which in turn implies the logarithmic growth in time of the bipartite entanglement entropy.}
    \label{fig:schematic-summary}
\end{figure}

We discover that the eigenstates forming these special quartets have a very specific structure in real as well as in Hilbert space. Note that, the MBL eigenstates have area-law entanglement~\cite{bauer2013area} notwithstanding the multifractality and rare resonances in Hilbert space~\cite{tikhonov2021eigenstate,roy2022hilbert,biroli2023largedeviation}. It is therefore a reasonable starting point to consider that a typical MBL eigenstate has a well-defined localisation centre in Hilbert space. This is nothing but the infinite disorder eigenstate to which the original eigenstate is connected via a finite-depth local unitary operator~\cite{bauer2013area,serbyn2013local,huse2014phenomenology,roy2022hilbert} and which is manifestly a product state. We find that the localisation centres of the MBL eigenstates in the special quartets have a specific structure -- the four states are made up of combinations of two different states in $A$ and two different in $B$. 
As such, for any eigenstate in such a quartet, there are two others which differ from the first one only in one subsystem whereas the remaining fourth one differs from the first one in both the subsystems. This is shown schematically in Fig.~\ref{fig:schematic-summary}(a). Denoting Hilbert space dimension of subsystem $A (B)$ as $N_{{\cal H}_{A(B)}}$, the number of choosing two different states in $A (B)$ is $O(N_{\mathcal{H}_{A(B)}}^2)$ such that total number of such quartets is $O(N_{\mathcal{H}_{A}}^2N_{\mathcal{H}_{B}}^2) \sim O(\nh^2)$. This explains the number of such special quartets.

This real-spatial structure of the special dominant quartets provides us with an organising principle for grouping them. Since the interactions in the system are short-ranged, the entanglement can spread only locally. This suggests grouping the quartets in terms of the distance, $r$, of the nearest sites from the bipartition (between $A$ and $B$) that differ amongst the states in the quartets. Via numerical exact diagonalisation, we obtain the frequencies associated these special dominant quartets, which as mentioned earlier are sums and differences of the eigenvalues of the four states involved in a quartet. Interestingly, we find that the distributions of the frequencies has a distribution which exhibits a scaling form -- the distributions for different $r$ can be collapsed onto a common curve when the frequencies are scaled with a characteristic frequency $\omega_\ast(r)$ which in turn decays exponentially with $r$. We show that the interplay of this hierarchy in the energyscales $\omega_\ast(r)$ and the number of such special dominant quartets with $r$ leads to the emergence of the power-law in $\omega$ decay of $\puo(\omega)$ which in turn implies the logarithmic growth of the second R\'enyi entropy of entanglement in time.

Resolving the dynamics in $r$ and the associated energyscales also provides a window into the spacetime picture of the entanglement growth. In particular, from the distribution of the frequencies for the quartets with a given $r$, we find that the entanglement growth at time $t$ is dominated by the special quartets with $r(t)\sim \ln t$. This leads to the notion of a logarithmically spreading `entanglement wavefront' as schematically shown in Fig.~\ref{fig:schematic-summary}(b). Since the degrees of freedom within the two wavefronts are strongly entangled, it again implies the logarithmic growth of bipartite entanglement entropy.

We finally discuss how the eigenstate and spectral correlations also lead to the volume-law saturation of the entanglement entropy in the MBL phase and an area-law in the Anderson localised case.

\subsection{Organisation of the paper}
The rest of the paper is organised as follows. In Sec.~\ref{sec:framework} we lay out the general framework and show how the dynamics of purity is encoded in the four-point eigenstate and spectral correlations. In particular in Sec.~\ref{sec:psi0-avg} we show how random product states as initial states can be averaged over and in Sec.~\ref{sec:dyn-sat} we discuss how to separate the dynamical and infinite-time components of the purity. The connection between eigenstate correlations and the operator entanglement of the time-evolution operator is described in Sec.~\ref{sec:opee}. In Sec.~\ref{sec:model}, we present some numerical results for a Floquet, disordered spin-1/2 chain which serves as the test bed for our theory. Section~\ref{sec:anatomy} consists of a detailed analysis of the eigenstate correlations that contribute to the dynamics of purity and forms the backbone of the paper. In Sec.~\ref{sec:special-quart} we identify the special quartets which dominate the dynamics of purity and their structure, both in real and Fock space. This structure leads to a hierarchy of frequency and timescales which we is discussed in Sec.~\ref{sec:hierarchy}, and from which the logarithmic growth of entanglement emerges automatically as discussed in Sec.~\ref{sec:log-growth}. The spacetime picture of entanglement, as emerges from the hierarchy of frequency and timescales, constitutes Sec.~\ref{sec:spcaetime}. In
Sec.~\ref{sec:ellbit}, we discuss how our work is related to and goes beyond the $\ell$-bit picture. We briefly discuss the infinite-time saturation of the purity in Sec.~\ref{sec:inftime} before closing with a summary and outlook in Sec.~\ref{sec:conclusion}.

\section{Dynamics of purity and eigenstate correlations \label{sec:framework}}
In this section, we lay out the basic framework and show the how the dynamics of purity is encoded in dynamical eigenstate correlations. In particular, we show that averaging over initial states which are random product states between the subsystems, leads to an expression for the purity purely in terms of eigenstates correlations which involves sets of four eigesntates. We also discuss the subset of eigenstate correlations which encode the infinite-time purity. It is worth mentioning here that the framework discussed in this section is completely general and is equally valid for an ergodic or an MBL system.

\subsection{Setting and definitions \label{sec:def}}
We start by describing the basic setting and establishing some necessary notation. Consider a quantum system with a bipartition into two subsystems, $A$, and its complement, $B$ (see Fig.~\ref{fig:schematic-summary}). For a given state $\ket{\psi}$ of the system, the second R\'enyi entropy of entanglement between $A$ and $B$ is given by $S_2^{AB} = -\ln \pu$ where the bipartite purity, $\pu$, is defined as 
\eq{
    \pu = {\rm Tr_A}[\rho_A^2];\quad \rho_A = {\rm Tr}_B[\ket{\psi}\bra{\psi}]\,,
    \label{eq:purity}
}
where $\rho_A$ is the reduced density matrix of subsystem $A$ obtained by a partially trace of the density matrix of the full system over the degrees of freedom in subsystem $B$.

The Hilbert space of the system is made up of a tensor product of the Hilbert spaces of $A$ and $B$, ${\cal H} = {\cal H}_A\otimes {\cal H}_B$. Let us denote a set of orthonormal basis state of ${\cal H}_A$ as $\{\ket{i_A}\}$ and similarly of ${\cal H}_B$ as $\{\ket{i_B}\}$ such that any basis state of ${\cal H}$ is given by $\ket{i_A,i_B}\equiv \ket{i_A}\otimes\ket{i_B}$. With this notation, the purity in Eq.~\ref{eq:purity} can be expressed as 
\eq{
    \pu = \sum_{\substack{i_A,i_B,\\j_A,j_B}}\psi_{i_Ai_B}\psi_{i_Aj_B}^\ast\psi_{j_Ai_B}^\ast\psi_{j_Aj_B}\,,
    \label{eq:pu-psi}
}
where $\psi_{i_Ai_B}=\langle i_A,i_B\ket{\psi}$. Note that Eq.~\ref{eq:pu-psi} is manifestly invariant under arbitrary basis transformations within ${\cal H}_A$ and ${\cal H}_B$. Also note that, Eq.~\ref{eq:pu-psi} holds for a time-evolving state as well where the eigenstate amplitudes and hence the purity depend explicitly on time.

In order to study the dynamics of purity, we consider a periodically driven, or a Floquet system
where we denote the time-evolution operator over one period as $U_F$ which is also often referred to as the Floquet unitary. While all the results in this work hold for a system described by a time-independent Hamiltonian, we consider a Floquet system for two reasons. First, in the latter all (quasi)energies are statistically equivalent and the density of states is flat -- this removes the necessity to unfold the spectrum while studying dynamical eigenstate correlations. Second, Floquet systems do not allow for mobility edges which serves as a convenience as we prefer to not have results in the MBL regime potentially contamimated by them. We denote by $\{\ket{\alpha}\}$ and $\{\theta_\alpha\}$, the set of eigenstates and eigen-(quasi)energies of $U_F$, such that
\eq{
    U_F = \sum_{\alpha}e^{-i\theta_\alpha}\ket{\alpha}\bra{\alpha}\,,
    \label{eq:uf}
}
with $\alpha_{i_Ai_B} = \langle i_A,i_B\ket{\alpha}$. An initial state, $\ket{\psi_0}$ at time $t=0$, evolves in time with the unitary in Eq.~\ref{eq:uf} as $\ket{\psi(t)} = U_F^t\ket{\psi_0}$ such that the time-dependent amplitudes are given by 
\eq{
    \psi_{i_Ai_B}(t) = \sum_\alpha e^{-i \theta_\alpha t}\alpha_{i_Ai_B}\bra{\alpha}\ket{\psi_0}\,.
}
Using the above in Eq.~\ref{eq:pu-psi}, we obtain an expression for the purity at time $t$ as
\eq{
    \pu_{\psi_0}(t) = \sum_{\al,\be,\ga,\la}e^{-it\tabcd}V_{\al\be\ga\la}\mathcal{I}^{\al\be\ga\la}_{\psi_0}\,,
    \label{eq:pu-eigenstate-psi0}
}
where 
\eq{
    \tabcd = \theta_{\alpha}-\theta_\beta-\theta_\gamma+\theta_\lambda\,,
    \label{eq:theta-abcd}
}
are the different frequencies composed of sums and differences of the quasienergies that contribute to the dynamics, 
\eq{
    V_{\al\be\ga\la} = \sum_{\substack{i_A,i_B,\\ j_A,j_B}}\alpha_{i_Ai_B}\beta^\ast_{i_Aj_B}\gamma_{j_Ai_B}^\ast\lambda_{j_Aj_B}\,,
    \label{eq:V-abcd}
}
denotes the corresponding eigenvector correlations involving the four different eigenstates, and 
\eq{
    \mathcal{I}^{\al\be\ga\la}_{\psi_0}=\braket{\al}{\psi_0}\braket{\psi_0}{\beta}\braket{\psi_0}{\gamma}\braket{\la}{\psi_0}\,,
    \label{eq:I-abcd-psi0}
}
contains explicitly the information of the initial state. This explicit dependence on the initial state is also reflected in the subscript in $\pu_{\psi_0}(t)$ in Eq.~\ref{eq:pu-eigenstate-psi0}. It is important to note that $\Vabcd$ in Eq.~\ref{eq:V-abcd} is again invariant under arbitrary basis transformations within subsystems ${\cal H}_A$ and ${\cal H}_B$.
As a final piece notation, we introduce $\puo(\omega)$ denoting the dynamics of purity in the frequency domain. Formally the Fourier transform of Eq.~\ref{eq:pu-eigenstate-psi0}, it can be expressed as 
\eq{
    \puo_{\psi_0}(\omega) = \sum_{\al,\be,\ga,\la}\delta_{2\pi}(\tabcd-\omega)V_{\al\be\ga\la}\mathcal{I}^{\al\be\ga\la}_{\psi_0}\,.
    \label{eq:puo-eigenstate-psi0}
}
where the subscript of $2\pi$ in the Dirac-delta function denotes that $\tabcd=\omega$ mod $2\pi$ as we will be working with Floquet systems.
Note that the each term in the sum in Eq.~\ref{eq:pu-eigenstate-psi0} or in Eq.~\ref{eq:puo-eigenstate-psi0} is a correlation involving eight distinct eigenstate amplitudes, four from the eigenstate correlation in Eq.~\ref{eq:V-abcd} and four from the initial state dependence in Eq.~\ref{eq:I-abcd-psi0}.

\subsection{Purity averaged over initial states \label{sec:psi0-avg}}
While the results for purity in Eq.~\ref{eq:pu-eigenstate-psi0} and in Eq.~\ref{eq:puo-eigenstate-psi0} depend explicitly on the initial state, we next show that they can be averaged over a large family on initial state such that they depend only on eigenstate and spectral correlations. In particular, we consider random product states between $A$ and $B$ as initial states
\eq{
    \ket{\psi_0} = \ket{\psi_0^A}\otimes\ket{\psi_0^B}\,,
    \label{eq:psi0-product}
}
where $\ket*{\psi_0^{A}}$ is a normalised Haar random state in susbsystem $A$ and similarly in $B$.
The motivation behind this choice of initial states is twofold. First, the growth of entanglement starting from such states is quantitatively the same as the growth of the operator entanglement entropy of $U_F^t$~\cite{lezama2019powerlaw}, and hence quantifies entanglement growth in a manner which is not biased due to choices of initial states.
The second, and arguably more important in this case, point is that for such initial states, the quantity $\mathcal{I}^{\al\be\ga\la}_{\psi_0}$ (see Eq.~\ref{eq:I-abcd-psi0}), encoding the information of the initial state, can be straightforwardly averaged over all $\psi_0$ of the form in Eq.~\ref{eq:psi0-product}. Using the Haar random nature of $\ket*{\psi_0^{A}}$ and $\ket*{\psi_0^{B}}$, it can be shown that (see Appendix~\ref{app:psi0-avg})
\eq{
    \mathbb{E}\left[{\cal I}^{\alpha\beta\gamma\lambda}_{\psi_0}\right]_{\psi_0} = 
    \frac{1}{\nh^2}[\delta_{\al\be}\delta_{\ga\la}+\delta_{\al\ga}\delta_{\be\la}+\nonumber\\
    V_{\al\be\ga\la}^*+V_{\al\ga\be\la}^*]\,,
    \label{eq:I-abcd-avg}
}
where $\mathbb{E}[\cdots]_{\psi_0}$ denotes the average over intial states and $V_{\al\be\ga\la}$is given by Eq.~\ref{eq:V-abcd}. Defining the initial-state averaged purity as $\puo(\omega) \equiv \mathbb{E}[\puo_{\psi_0}(\omega)]_{\psi_0}$, an expression for it can be obtained using Eq.~\ref{eq:I-abcd-avg} in Eq.~\ref{eq:puo-eigenstate-psi0} as 
\eq{
\begin{split}
    \puo(\omega) = &\frac{1}{\nh^2}\sum_{\al,\be}(V_{\al\be\al\be}+V_{\al\al\be\be})\delta(\omega)+\\&\frac{1}{\nh^2}\sum_{\al,\be,\ga,\la}\delta(\ta_{\al\be\ga\la}-\omega)V_{\al\be\ga\la}(V_{\al\be\ga\la}^\ast+V_{\al\ga\be\la}^\ast)\,.
    \label{eq:puo-initial-avg}
\end{split}
}
An obvious but key point to note here is that the average over the initial states has yielded an expression for the dynamics of purity, \eqref{eq:puo-initial-avg}, that depends only on the eigenstate and spectral correlations.

\subsection{Dynamical and infinite-time purity \label{sec:dyn-sat}}
At this stage, it will be particularly useful to separate the dynamical part of $\puo(\omega)$ corresponding to $\omega\neq 0$ from the infinite-time saturation, corresponding to $t\to\infty$ or equivalently $\omega=0$ for a finite system. As such, we express $\puo(\omega)=\pudyn(\omega)+\pusat\delta(\omega)$. 

The infinite-time saturation of purity is given by 
\eq{
    \begin{split}
    \pusat = \frac{1}{\nh^2}[\sum_{\al,\be}(V_{\al\be\al\be}&+V_{\al\al\be\be})+\\\sum_{\substack{\al,\be,\ga,\la:\\\tabcd=0}}&V_{\al\be\ga\la}(V_{\al\be\ga\la}^\ast+V_{\al\ga\be\la}^\ast)]\,.
    \end{split}
    \label{eq:pu-sat}
}
We will return to a detailed analysis of the above expression in Sec.~\ref{sec:inftime} and discuss the scaling of $\pusat$ with system size and show it leads to a volume-law saturation of the bipartite entanglement entropy.

However, much of our focus in this work is on the dynamics of purity which is encoded in $\pudyn(\omega)$, given by  
\eq{
    \begin{split}
    \pudyn(\omega) = \frac{1}{\nh^2}\sum_{\substack{\al,\be,\ga,\la:\\\tabcd\neq0}}\delta(\ta_{\al\be\ga\la}-&\omega)V_{\al\be\ga\la}\times\\&(V_{\al\be\ga\la}^\ast+V_{\al\ga\be\la}^\ast)\,,
    \end{split}
    \label{eq:pu-dyn}
}
where it can be demonstrated that the second term in the brackets is negligibly small such that, 
\eq{
    \pudyn(\omega) = \frac{1}{\nh^2}\sum_{\substack{\al,\be,\ga,\la:\\\tabcd\neq0}}\delta(\ta_{\al\be\ga\la}-&\omega)|V_{\al\be\ga\la}|^2\,.
    \label{eq:pu-dyn-V2}
}
The expression in Eq.~\ref{eq:pu-dyn-V2} constitutes a key ingredient to our theory of entanglement growth and its right hand side is a central quantity of analysis, as will become clear in the subsequent sections. While details of the calculation and numerical evidence for ignoring the second term in Eq.~\ref{eq:pu-dyn} are presented in Appendix~\ref{app:pudyn}, we argue for it here on general grounds. 

Note that $\puo(\omega)$ satisfies a sum rule, namely $\int d\omega~\puo(\omega) = 1$, which is a result of the initial state being a product state between $A$ and $B$, such that $\pu(t=0)=1$.
At the same time, using the orthonormality of the eigenvectors, it can be shown that
\eq{
    \sum_{\alpha,\beta,\gamma,\lambda}|V_{\alpha\beta\gamma\lambda}|^2 = \nh^2\,; \sum_{\alpha,\beta,\gamma,\lambda}V_{\alpha\beta\gamma\lambda}V_{\alpha\gamma\beta\lambda}^\ast = \nh\,.
    \label{eq:Vabcd-sum-rule}
}
Interpreting $\puo(\omega)$ as a distribution of $\tabcd$ weighted by the respective $|V_{\alpha\beta\gamma\lambda}|^2+V_{\alpha\beta\gamma\lambda}V_{\alpha\gamma\beta\lambda}^\ast$, Eq.~\ref{eq:Vabcd-sum-rule} makes it clear that almost the entire weight of the distribution is carried by the first term and the contribution of the second term is negligibly small, $O(\nh^{-1})$. It is therefore, completely justfied to ignore the second term, doing precisely which yields Eq.~\ref{eq:pu-dyn-V2} from Eq.~\ref{eq:pu-dyn}.

\subsection{Operator entanglement entropy of time-evolution operator\label{sec:opee}}

It is also useful to note that the dynamics of bipartite entanglement averaged over the initial states, as in Sec.~\ref{sec:psi0-avg}, is directly related to the operator entanglement entropy~\cite{prosen2007operator,zhou2017operator} of the time-evolution operator $U(t) \equiv U_F^t$ with $U_F$ given in Eq.~\ref{eq:uf}. To see this, note that the time-evolution operator, just like any other operator, can be written as a vector in a doubled Hilbert space, of dimension $\nh^2$), which is the Hilbert space of all operators defined for the system. The inner product between two operators, $\cal{X}$ and $\cal{Y}$, is defined as 
\eq{
({\cal{X}|\cal{Y}}) = \frac{1}{\nh}{\rm Tr}[{\cal X}^\dagger {\cal Y}]\,,
\label{eq:op-ip}
}
where the notation $|\cdots)$ denotes an operator written as a vector. The doubled Hilbert space can also be written as a direct product of the doubled Hilbert spaces of the two subsystems and each of them are spanned by the set of operators $\{{\cal O}_A\}$ and $\{{\cal O}_B\}$ which have support only on their respective subsystems and are mutually orthonormal under the definition in Eq.~\ref{eq:op-ip}. Using this notation, the time-evolution operator can be expressed as 
\eq{
|U(t)) = \sum_{{\cal O}_A,{\cal O}_B}
u_{{\cal O}_A,{\cal O}_B}(t)|{\cal O}_A)\otimes|{\cal O}_B)\,,
}
where $u_{{\cal O}_A,{\cal O}_B}(t)=({\cal O}_A\otimes{\cal O}_B|U(t))$.
The bipartite operator purity of the time-evolution operator is then 
\eq{
{\cal P}_{U(t)}^{AB} =  \sum_{\substack{{\cal O}_A,{\cal O}_B,\\{\cal O}^\prime_A,{\cal O}^\prime_B}}
u_{{\cal O}_A{\cal O}_B}
u^\ast_{{\cal O}^\prime_A{\cal O}_B}
u^\ast_{{\cal O}_A{\cal O}^\prime_B}
u_{{\cal O}^\prime_A{\cal O}^\prime_B}\,,
\label{eq:op-pu}
}
where the explicit dependence of $u_{{\cal O}_A{\cal O}_B}$ on $t$ as been suppressed for brevity. 

While the operator purity in Eq.~\ref{eq:op-pu} as well as the eigenstate correlations, $\Vabcd$ in Eq.~\ref{eq:V-abcd} are basis independent, the explicit relation between the two is most conveniently seen as follows. For any set of orthonormal basis states $\{\ket{i_A}\}$ of ${\cal H}_A$, one can construct a set of operators ${\cal A}_{i_Aj_A}=\sqrt{N_{{\cal H}_A}}\ket{j_A}\bra{i_A}$ which form a set of orthonormal basis for the operator Hilbert space of subsystem $A$; and similarly for $B$. This particularly choice of ${\cal O}_{A/B}$ leads to 
\eq{
u^{{\phantom\dagger}}_{{\cal A}_{i_Aj_A}{\cal B}_{i_Bj_B}} (t)= \frac{1}{\sqrt{\nh}}\sum_{\alpha}e^{-i\theta_\alpha t}\alpha_{i_Ai_B}\alpha_{j_Aj_B}^\ast\,,
}
which when used in Eq.~\ref{eq:op-pu} directly leads to 
\eq{
{\cal P}_{U(t)}^{AB} = \frac{1}{\nh^2}\sum_{\alpha,\beta,\gamma,\lambda}|\Vabcd|^2e^{-it\tabcd}\,,
}
which is nothing but the bipartite purity of time-evolving states initialised as product states, Eq.~\ref{eq:psi0-product}, upon averaging over such initial states. 
This is particularly useful because the rest of what follows can be understood as a unified picture for entanglement growth starting from typical product states as well as entanglement growth of the time-evolution operator.

\section{Disordered Floquet spin-1/2 chain as a model \label{sec:model}}
While the entire framework relating the dynamics of purity to eigenstate correlations described in the previous section was completely general, our main focus on strongly disordered systems in the MBL regime. 
In order to orient ourselves in that direction, and also as a concrete setting for demonstrating our theory, we employ a disordered, Floquet spin-1/2 chain. The Floquet unitary, $U_F$, is given by~\cite{zhang2016floquet}
\eq{
U_F = \exp[-i\tau H_X]\,\exp[-i\tau H_Z]\,,
\label{eq:UF}
}
where
\eq{
    \begin{split}
        H_X &=g\Gamma\sum_{i=1}^L \sigma^x_i\,,\\
        H_Z &= \sum_{i=1}^L [\sigma^z_i\sigma^z_{i+1} + (h+g\sqrt{1-\Gamma^2}\epsilon_i)\sigma^z_i]\,,
    \end{split}
    \label{eq:HXHZ}
}
where $\{\sigma_i^\mu\}$ is the set of Pauli matrices representing the spins-1/2 and $\epsilon_i\sim \mathcal{N}(0,1)$ are Gaussian random numbers with zero mean and unit standard deviation. Following Ref.~\cite{zhang2016floquet} we take $g = 0.9045$, $h = 0.809$, and $\tau = 0.8$. For these parameters, there is a putative many-body localisation transition at $\Gamma_c\approx 0.3$ with the model in an ergodic phase for $\Gamma>\Gamma_c$ and in a MBL regime for $\Gamma<\Gamma_c$. It is the latter that we are interested in and hence, all our numerical results are for $\Gamma=0.1$ and $0.15$, two representative values in the MBL regime.

\begin{figure}
\includegraphics[width=\linewidth]{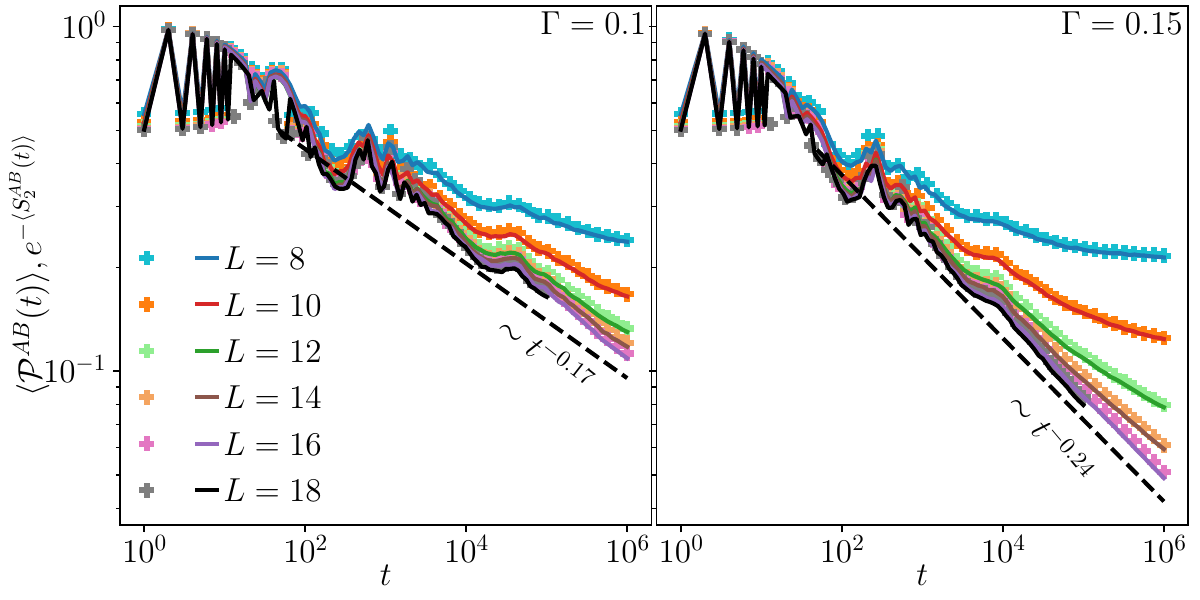}
\caption{The decay (growth) of purity ($2^{\rm nd}$-R\'enyi entropy) with time in the MBL phase showing a power-law decay (logarithmic growth). In each panel, the data points denote the average purity, $\langle\pu(t)\rangle$, whereas the lines denote the exponential of the averaged $2^{\rm nd}$-R\'enyi entropy, $e^{-\langle S_2^{AB}(t)\rangle}$. The data shows that $\langle\pu(t)\rangle\approx e^{-\langle S_2^{AB}(t)\rangle}$ and hence it sufficient to focus on the former. The dashed lines are guides to the eye for the power-law decay of $\pu(t)$ with $t$. The two panels correspond to two different values of $\Gamma$ in the MBL regime and different colours denote different system sizes $L$.}
\label{fig:Pt-S2t}
\end{figure}

In Fig.~\ref{fig:Pt-S2t}, we show the results for $\expval{\pu(t)}$ and $\expval{S_2^{AB}(t)}$ as a function of time $t$, where the $\expval{\cdots}$ denotes the average over disorder realisations. The first thing to note from the data is that $\expval{\pu(t)}$ is extremely close to $\exp[-\expval{S_2^{AB}(t)}]$. That the quenched and annealed averages are so close to each other suggests that the distribution of the purity remains extremely narrow at all times. However the key takeaway from this is that it is indeed sufficient to $\pu(t)$ or $\puo(\omega)$, and their averages over disorder realisations to understand the dynamics of $\expval{S_2^{AB}(t)}$.

The data shown in Fig.~\ref{fig:Pt-S2t} suggests that with increasing system size $L$, $\expval{\pu(t)}$ as a function of $t$ on logarithmic axes falls better and better onto a straight line. As such the data is best described by a power-law of the form
\eq{
\expval{\pu(t)}\sim t^{-a}\,,
\label{eq:Pt-power-law}
}
or equivalently, $\expval{S^{AB}_2(t)}\sim a\ln t$ with  $0<a<1$. It is also consistent with expectation that the value of $a$ is smaller for $\Gamma=0.1$ compared to that for $\Gamma = 0.15$, as the purity is expected to decay slower in the former case by virtue of it being deeper in the MBL regime.

\begin{figure}
\includegraphics[width=\linewidth]{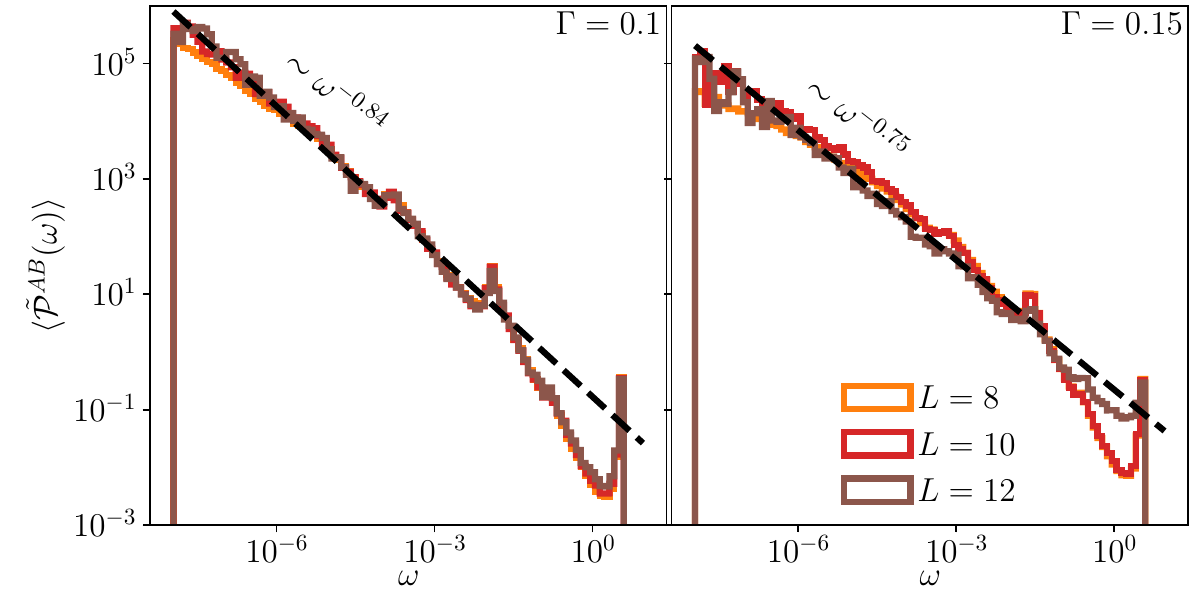}
\caption{The subsystem purity in the frequency domain, $\langle \puo(\omega)\rangle$ as a function of $\omega$ for the same two values of $\Gamma$ as in Fig.~\ref{fig:Pt-S2t}, for different $L$. The data shows a power-law decay in $\omega$ with the exponent concomitant with the power-law decay of $\langle\pu(t)\rangle$ with $t$. The dashed lines denote the fits used to extract the exponents.}
\label{fig:Pomega}
\end{figure}

Turning towards numerical results in the frequency domain, Fig.~\ref{fig:Pomega} shows $\langle\puo(\omega)\rangle$ evaluated explicitly~\footnote{Evaluating $\Vabcd$ and $\tabcd$ for all the $O(\nh^4)$ quartets of eigenstates for $L>8$ is computationally unfeasible. We therefore randomly choose $10^7$ quartets of eigenstates for each disorder realisation and average the results over them as well as over around $10^4$ disorder realisations. We however checked that for $L=8$, the results obtained this way match the ones obtained by considering all the quartets.} using Eq.~\ref{eq:puo-initial-avg}. The data is again best described by a power-law decay in $\omega$,
\eq{
\expval*{\puo(\omega)}\sim \omega^{-b}\,,
\label{eq:P-omega-powerlaw}
}
particularly for $\omega\ll 1$ corresponding to long-time dynamics. It is useful to point out that $b\approx 1-a$, although it is expected on general grounds as $\puo(\omega)$ is essentially the Fourier transform of $\pu(t)$. Concomitantly, $b$ increases on moving deeper into the MBL phase. 
This is an interesting point of distinction between the higher-point correlations of the form in Eq.~\ref{eq:pu-dyn-V2} and correlations involving only two eigenstates. Albeit basis dependent, the latter also exhibits a power-law dependence in $\omega$ with the corresponding exponent becoming smaller on moving deeper into the MBL phase~\cite{tikhonov2021eigenstate}. The fact that this can be explained via rare resonances between pairs of eigenstates~\cite{tikhonov2021eigenstate} and is in complete contrast to the behaviour of the exponent $b$ in Eq.~\ref{eq:P-omega-powerlaw} suggests that the underlying physics at the heart of $\puo(\omega)$ is different from and goes beyond pairwise resonances between two eigenstates.

Understanding microscopic origins of the power-law in $\omega$ behaviour of $\puo(\omega)$, and possible subleading corrections will be the central focus of the rest of the paper.
However, the subleading corrections to the behaviour in Eq.~\ref{eq:Pt-power-law} and in Eq.~\ref{eq:P-omega-powerlaw}, if present, are unlikely to be revealed in the data shown in Fig.~\ref{fig:Pt-S2t} and Fig.~\ref{fig:Pomega} due to limited system sizes and timescales accessible in numerical calculations.

\section{Anatomy of eigenstate correlations \label{sec:anatomy}}

The key takeaway from the general framework laid out in Sec.~\ref{sec:framework} was the explicit connection between the dynamics of purity and the eigenstate correlations, as embodied in, for example Eq.~\ref{eq:pu-dyn-V2}. It is therefore clear that to develop a theoretical understanding of the dynamical behaviour, as exemplified in Sec.~\ref{sec:model}, a detailed understanding of the anatomy of the eigenstate correlations, $|\Vabcd|^2$, and the associated spectral correlations is important. This constitutes the topic of this section.

\subsection{Special quartets which dominate dynamics of purity \label{sec:special-quart}}
\begin{figure}
\includegraphics[width=\linewidth]{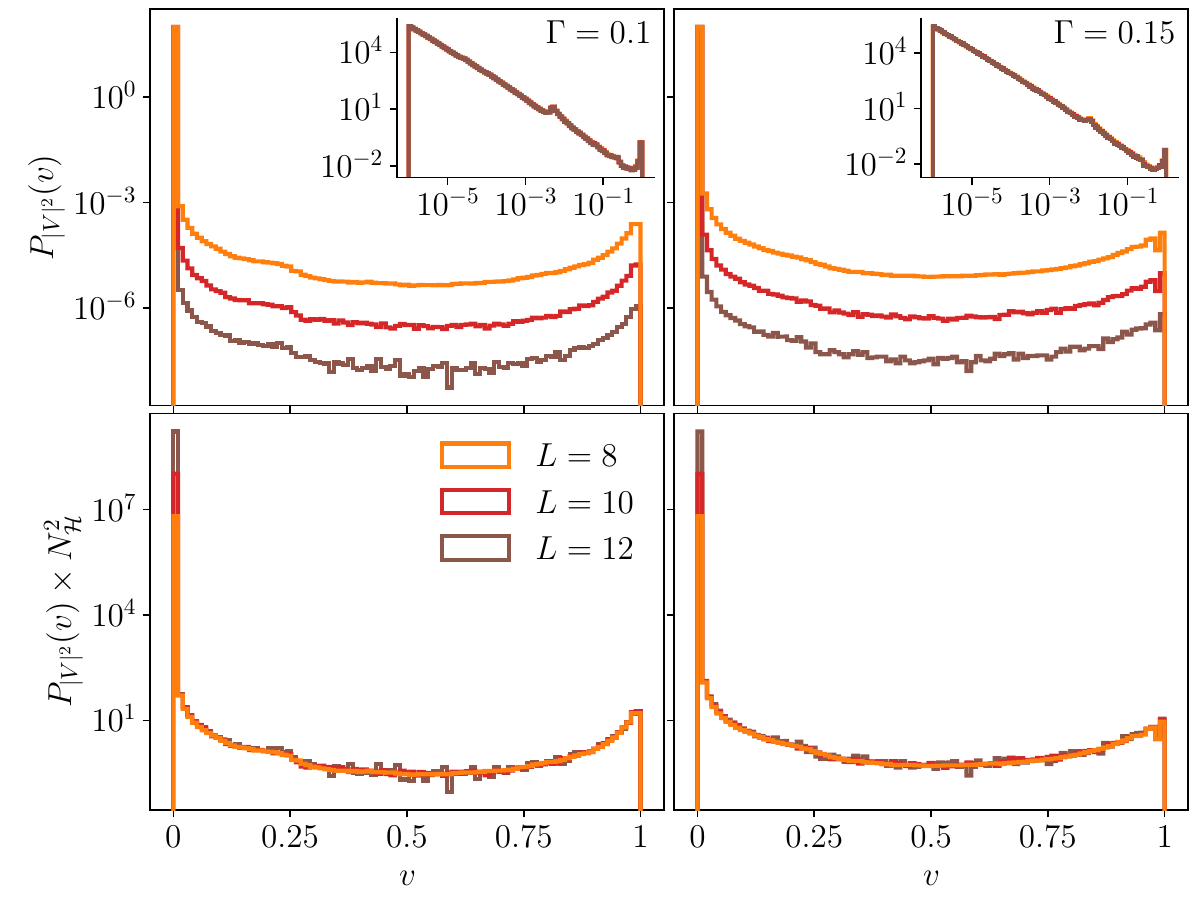}
\caption{The probability distribution $P_{|V|^2}$ (defined in Eq.~\ref{eq:P-Vsq}) of  $|\Vabcd|^2$ over all quartets for which $\tabcd\neq 0$. The top panels show the bare distributions whereas the bottom panels show the distributions scaled by $\nh^2$. The insets in the top panels show the raw distributions but with logarithmic binning on logarithmic scales. The left and right panels correspond to $\Gamma=0.1$ and 0.15 respectively, and different colours denote different $L$.
}
\label{fig:Vdist}
\end{figure}

To get a basic idea of the eigenstate correlations, $|\Vabcd|^2$, we begin by studying their distributions. Since we are interested in the dynamics of purity, as encoded in $\pudyn(\omega)$ , we focus on the distribution $|\Vabcd|^2$ over quartets of eigenstates for which $\tabcd\neq 0$ (see Eq.~\ref{eq:pu-dyn-V2}). Denoted by $P_{|V|^2}$, the distribution is defined as 
\eq{
P_{|V|^2}(v)=
\frac{1}{{\cal N}_Q}\sum_{\substack{\al,\be,\ga,\la:\\\tabcd\neq0}}\delta(|V_{\al\be\ga\la}|^2-v)\,,
\label{eq:P-Vsq}
}
where the normalisation, ${\cal N}_Q=\nh^4 -2\nh^2+\nh\approx\nh^4$, is simply the number of quartets that satisfy $\tabcd\neq 0$.
The results are shown in Fig.~\ref{fig:Vdist} for the same two values of $\Gamma$ as in Fig.~\ref{fig:Pt-S2t} and Fig.~\ref{fig:Pomega}.

The important feature that emerges out of the results is that for an overwhelmingly large fraction, in fact close to unity, of the quartets of eigenstates, $|\Vabcd|^2$ is vanishingly small. This is reflected in the large peak in $P_{|V|^2}(v)$ near $v\approx 0^+$, and that the height of the peak is seemingly converged with system size (top panels). This is, in fact, better understood from the insets where the distributions are shown on logarithmic scales such that regime of small values $|\Vabcd|^2\ll 1$ is magnified.  On the other hand, the fact that $P_{|V|^2}(v)$ for $v\sim O(1)$ collapses for different $L$ when scaled by $\nh^{-2}$ (bottom panels). This suggests that for a number of quartets which scales as $\nh^2$ (or equivalently whose fraction scales as $\nh^{-2}$), $|\Vabcd|^2 \sim O(1)$.
This observation raises two questions, (i) what is the distinguishing feature of these special quartets of states with $|\Vabcd|^2 \sim O(1)$, and (ii) do these special quartets dominate the dynamics of purity such that it is sufficient to consider only such quartets in Eq.~\ref{eq:pu-dyn-V2}.

\begin{figure}
\includegraphics[width=\linewidth]{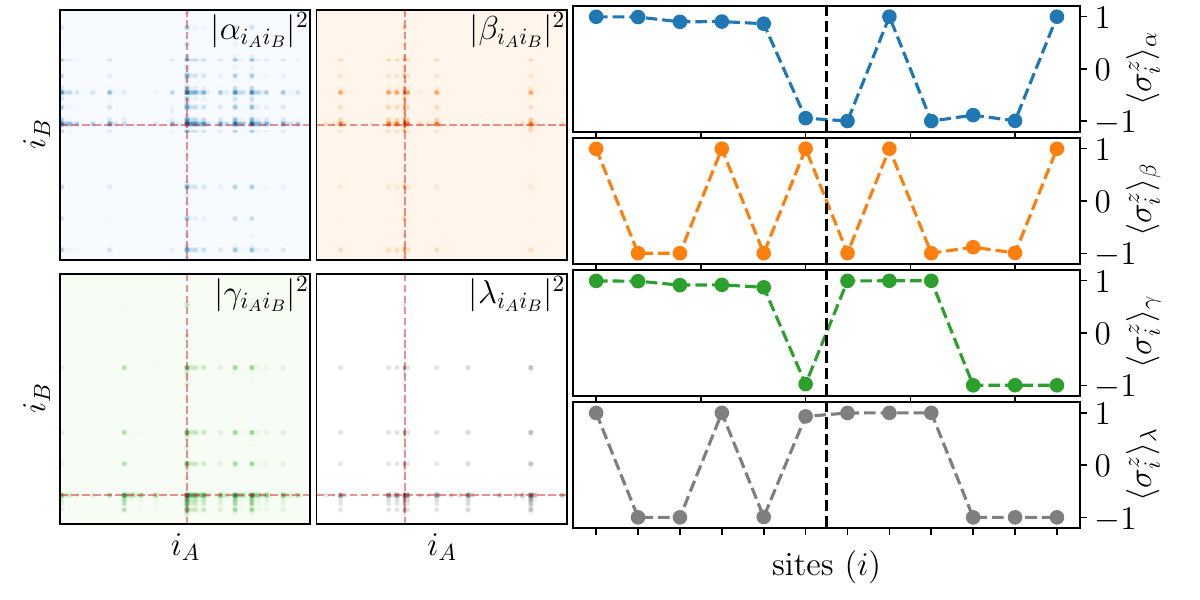}
\caption{Representative example of the structure of the eigenstates forming the special dominant quartets on the Fock space (left two columns) and real space (right column). 
The left two columns show the weights of the eigenstates as a colourmap on the Fock space of $\sigma^z$-product states with each panel corresponding to a particular eigenstate. Note that the localisation centres in Fock space (denoted by the red dashed lines) follow the pattern described in Eq.~\ref{eq:loc-pattern}. The right column shows the $\expval{\sigma^z_i}$ profiles along the chain for the four eigenstates. Note that the (dis)similarities in the profiles in the two subsystems are consistent with the localisation centres of the eigenstates. The data is for $L=12$ with $\Gamma=0.1$.}
\label{fig:Sq-struture}
\end{figure}

We start by investigating the first question. To understand the structure of these special quartets of eigenstates, we study their amplitudes in Fock space. Note that even though the MBL eigenstates exhibit (multi)fractal statistics in any basis which is a tensor product of local (in real space) bases, they can typically be associated to a localisation centre in an appropriate such basis~\cite{detomasi2020rare,roy2021fockspace}. For the model in Eq.~\ref{eq:HXHZ}, the $\sigma^z$-product state basis is the appropriate basis as the disordered fields in $H_Z$ couple to the $\sigma^z_i$ operators. The localisation centres are then nothing but the $\sigma^z$-product states to which the MBL eigenstates are analytically connected to via finite-depth local unitary circuit~\cite{bauer2013area,roy2022hilbert}. We denote the localisation centre of an eigenstate, say $\ket{\alpha}$, by $(i_A^\alpha,i_B^\alpha)$. Using this notation, we find that for any typical quartet of eigenstates with $|\Vabcd|^2\sim O(1)$ and $\tabcd\neq0$, the localisation centres satisfy the condition
\eq{
\begin{split}
i_A^\alpha = i_A^\gamma,~i_A^\beta = i_A^\lambda~~{\rm with}~i_A^\alpha\neq i_A^\beta\,,\\
i_B^\alpha = i_B^\beta,~i_B^\gamma = i_B^\lambda~~{\rm with}~i_B^\alpha\neq i_B^\gamma\,,\\
\end{split}
\label{eq:loc-pattern}
}
or the combination equivalent to a permutation of the indices in Eq.~\ref{eq:loc-pattern} with $\beta \leftrightarrow \gamma$.
The condition in Eq.~\ref{eq:loc-pattern} implies that the localisation centres of $\ket{\alpha}$ and $\ket{\gamma}$ correspond to the same $\sigma^z$-configuration in subsystem $A$ and similarly for $\ket{\beta}$ and $\ket{\lambda}$; however, the two localisation centres are necessarily different. On the other hand, it is $\ket{\alpha}$ and $\ket{\beta}$ whose localisation centres correspond to the same $\sigma^z$-configuration in subsystem $B$ and similarly for $\ket{\gamma}$ and $\ket{\lambda}$, with the two localisation centres again being different. Evidence for this structure provided in Fig.~\ref{fig:Sq-struture}. In the left two columns, we show the eigenstate intensities, $|\alpha_{i_Ai_B}|^2$, (and similarly for the other three eigenstates) as colourmap in the $(i_A,i_B)$ plane. The localisation centres, also marked by the red dashed lines, are indicated by the points of highest intensity in each of the panels. Note that, it is clear from the positions of the localisation centres that they do indeed satisfy Eq.~\ref{eq:loc-pattern}. As a matter of notation, hereafter we will denote by $\alpha,\beta,\gamma,\lambda\in\square$ any such special quartet of eigenstates.

This specific structure of the eigenstates also manifest themselves in the real-space profile of the $\sigma^z$-expectation values, $\expval{\sigma^z_i}_\alpha\equiv \expval{\alpha|\sigma^z_i|\alpha}$. A characteristic signature of the MBL regime is that $\sigma^z$-expectation values are close to $\pm 1$ and not exponentially small in $L$ as they would be in an ergodic phase~\cite{luitz2016long}. The condition in Eq.~\ref{eq:loc-pattern} would suggest that
\eq{
\begin{split}
&~~~~~~\expval{\sigma^z_i}_\alpha = \expval{\sigma^z_i}_\gamma\,,~ \expval{\sigma^z_i}_\beta = \expval{\sigma^z_i}_\lambda\,~\forall i\in A\,,\\
&~~~~~~\expval{\sigma^z_i}_\alpha = \expval{\sigma^z_i}_\beta\,,~ \expval{\sigma^z_i}_\gamma = \expval{\sigma^z_i}_\lambda\,~\forall i\in B\,,\\
&\text{along with}\\
&~~~~~~\exists\,i\in A\,:\, \expval{\sigma^z_i}_\alpha \neq\expval{\sigma^z_i}_\beta\,,\\
&~~~~~~\exists\,i\in B\,:\, \expval{\sigma^z_i}_\alpha \neq\expval{\sigma^z_i}_\gamma\,.
\end{split}
\label{eq:loc-pattern-real}
}
The right column in Fig.~\ref{fig:Sq-struture} shows evidence for Eq.~\ref{eq:loc-pattern-real}.

The picture that therefore emerges for the structure of quartets of eigenstates with $|\Vabcd|^2\sim O(1)$ is as follows. The four eigenstates in the quartet are such that there localisation centres are made of two distinct $\sigma^z$-configurations in subsystem $A$ and two in $B$. In other words, the four distinct combinations of these states form the localisation centres of the four eigenstates. In fact, this picture straightforwardly explains the number of quartets with $|\Vabcd|^2\sim O(1)$. For a given $\ket{\alpha}$, the number of states $\ket{\beta}$ which has the same localisation centre in $B$ but different in $A$ is $N_{{\cal H}_A}-1$. Similarly, the number of states $\ket{\gamma}$ which has the same localisation centre as $\ket{\alpha}$ in $A$ but different in $B$ is $N_{{\cal H}_B}-1$. Fixing $\ket{\beta}$ and $\ket{\gamma}$ automatically fixes the state $\ket{\lambda}$, see Eq.~\ref{eq:loc-pattern} or Fig.~\ref{fig:Sq-struture}. Since there are $\nh$ choices of $\ket{\alpha}$ itself, the total number of quartets which satisfy Eq.~\ref{eq:loc-pattern}, which we denote by ${\cal N}_\square$ is therefore $\nh (N_{{\cal H}_A}-1)(N_{{\cal H}_B}-1)$. As the subsystems $A$ and $B$ correspond to left and right halves of the entire system, we have ${\cal N}_{\square} =\nh (\sqrt{\nh}-1)^2\approx \nh^2$ or equivalently ${\cal N}_{\square}/{\cal N}_{Q}\sim \nh^{-2}$. This immediately explains why a fraction $\sim \nh^{-2}$ of the quartets have $|\Vabcd|^2\sim O(1)$ (see Fig.~\ref{fig:Vdist}).
It is worth clarifying and reiterating a couple of points here. First, our focus is deep in the MBL phase where associating an eigenstate to a localisation centre is well-defined despite the multifractality of the eigenstate. Second, the theoretical picture that emerges in the following sections does not depend on an eigenstate being sharply localised in the Hilbert space. Instead, the identification should be thought of as a starting point for constructing the theory\footnote{In fact, the existence and the structure of the special quartets are generic to MBL systems; while the results in Fig.~\ref{fig:Vdist} and Fig.~\ref{fig:Sq-struture} are for the Floquet system in Eq.~\ref{eq:UF}, an identical structure emerges also for time-independent MBL Hamiltonians as shown in Appendix~\ref{app:ham}.}.

\begin{figure}
\includegraphics[width=\linewidth]{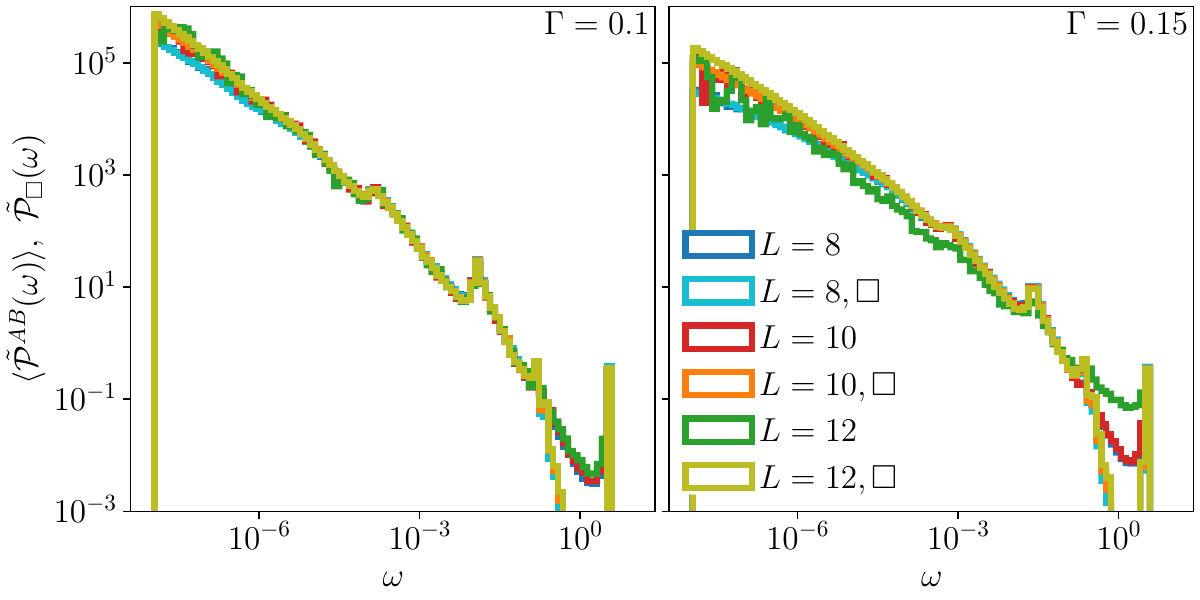}
\caption{The data for $\pusq(\omega)$, defined in Eq.~\ref{eq:P-sq}, as a function of $\omega$ is indistinguishable from that of the spectral correlations of the special quartets, $\langle\puo(\omega)\rangle$. This shows that the dynamics of purity is indeed dominated entirely by the spectral correlations within these quartets. The colours (see legend) labelled with $\square$ correspond to $\pusq(\omega)$ whereas the others to $\langle\puo(\omega)\rangle$, for different $L$. The left and right panels correspond respectively to $\Gamma=0.1$ and $\Gamma=0.15$. } 
\label{fig:pomega-q}
\end{figure}

We next turn to the second question raised above, namely do the special quartets with $|\Vabcd|^2\sim O(1)$ dominate the dynamics of purity. In order to ascertain that, we define 
\eq{
\pusq(\omega) = \frac{1}{\nh^2}\sum_{\alpha,\beta,\gamma,\lambda \in \square}\delta(\tabcd-\omega)\,,
\label{eq:P-sq}
}
where we restrict the sum in Eq.~\ref{eq:pu-dyn-V2} to only the special quartets, and also make the approximation that their $O(1)$ values of $|\Vabcd|^2$ can be set to 1.
The results for $\pusq(\omega)$ are shown in Fig.~\ref{fig:pomega-q} where it is clear that they are virtually indistinguishable from the full $\puo(\omega)$. This provides compelling evidence for the fact that the dynamics of purity is governed by spectral correlations (again four-point) between eigenstates forming the special quartets such that,
\eq{
\langle\puo(\omega)\rangle \approx \pusq(\omega)\,.
\label{eq:pu-pusq}
}
Understanding the power-law in $\omega$ of $\puo(\omega)$ then reduces to understanding these spectral correlations which we discuss next.

\subsection{Hierarchy and distribution of energyscales \label{sec:hierarchy}}

In order to understand the spectral correlations in Eq.~\ref{eq:P-sq}, it will be useful to organise the special dominant quartets using some guiding principles. 
The first of them is a consequence of
the locality of the interactions which results in the entanglement between $A$ and $B$ growing from the bipartition. In other words, sites in $A$ and $B$ which are closer to the bipartition will get entangled earlier than those further away. This suggests that the quartets should be classified based on a real-space distance measured from the bipartition. 
The second guiding principle is provided by the picture that the effective interactions between degrees of freedom in any effective theory must decay with the distance between them, for instance the decay is proposed to be exponential in the $\ell$-bit theory~\cite{serbyn2013local,huse2014phenomenology}..

\begin{figure}
\includegraphics[width=\linewidth]{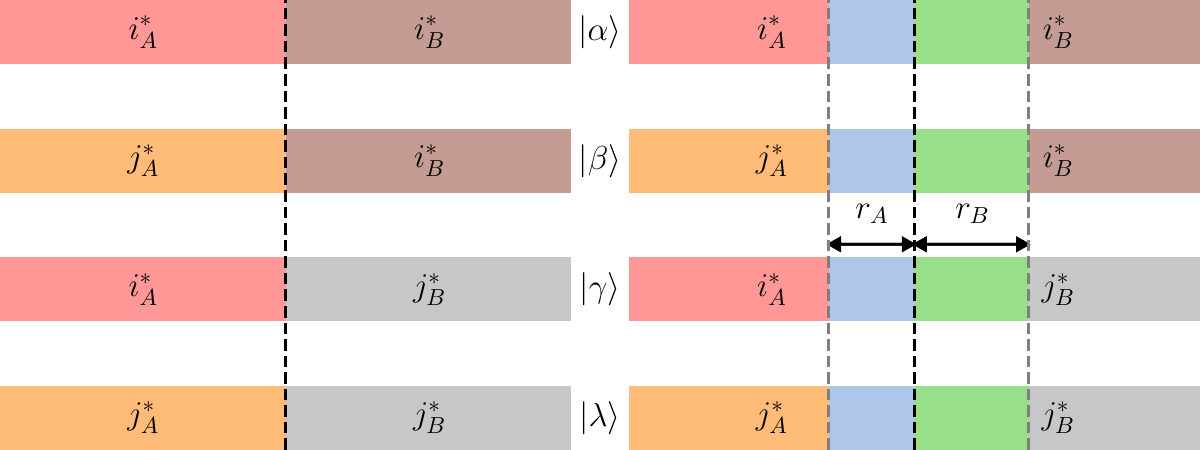}
\caption{Visualising the states forming the dominant quartets. Left: The localisation centres of the four eigenstates are made up of two different states in $A$, $\ket{i_A^\ast}$ and $\ket{j_A^\ast}$ (as indicated by their different colours), and two in $B$, $\ket{i_B^\ast}$ and $\ket{j_B^\ast}$. Right: Identifying the distances $r_{A(B)}$ a given such quartet; $r_{A(B)}$ denotes the distance from the bipartition of the first site that is different between $\ket{i_{A(B)}^\ast}$ and $\ket{j_{A(B)}^\ast}$. As such, the configuration on all the sites within a distance $r_{A(B)}$ from the bipartition are the same between $\ket{i_{A(B)}^\ast}$ and $\ket{j_{A(B)}^\ast}$ as indicated by the same colour.}
\label{fig:Sq-cartoon}
\end{figure}

Using this picture as the starting point, we organise the quartets as follows. Consider the localisation centres of the four eigenstates in a quartet to be made out of $\ket{i_A^\ast}$ and $\ket{j_A^\ast}$ in $A$ with $i_A^\ast\neq j_A^\ast$, and $\ket{i_B^\ast}$ and $\ket{j_B^\ast}$ in $B$ with $i_B^\ast\neq j_B^\ast$, see Fig.~\ref{fig:Sq-cartoon} (left panel) for a visual schematic. 
Note that $\tabcd$, in fact, is a difference of the differences of quasienergies $\theta_\alpha-\theta_\beta$ and $\theta_\gamma-\theta_\lambda$. Perturbatively, to  leading order in interactions, $\tabcd$ is therefore the difference,
\eq{
\tabcd\approx \Delta_{i_A^\ast j_A^\ast}^{i_B^\ast} - \Delta_{i_A^\ast j_A^\ast}^{j_B^\ast} + \cdots\,,
\label{eq:tabcd-diff-diff}
}
where $\Delta_{i_A^\ast j_A^\ast}^{i_B^\ast}$ is the difference between the energies of $\ket{i_A^\ast}$ and $\ket{j_A^\ast}$ due to interactions with $B$ in configuration $\ket{i_B^\ast}$ and similarly for $\Delta_{i_A^\ast j_A^\ast}^{j_B^\ast}$~\footnote{One can equivalently interpret $\tabcd$ as the difference  $\theta_\alpha-\theta_\gamma$ and $\theta_\beta-\theta_\lambda$ such that perturbatively it can be expressed as $\tabcd\approx \Delta_{i_B^\ast j_B^\ast}^{i_A^\ast} - \Delta_{i_B^\ast j_B^\ast}^{j_A^\ast} + \cdots$.}. This is where we use the second guiding principle above which suggests that the leading term in Eq.~\ref{eq:tabcd-diff-diff} will be dominated by the interaction between the site in $A$ and the site in $B$ which are the nearest to the bipartition but are different between $\ket{i_A^\ast}$ and $\ket{j_A^\ast}$ and between $\ket{i_B^\ast}$ and $\ket{j_B^\ast}$ respectively. We denote the distances of these sites from the bipartiton by $r_A$ and $r_B$ respectively, see Fig.~\ref{fig:Sq-cartoon} (right panel) for a visual schematic.
The upshot of the above is that given a special dominant quartet, $\alpha,\beta,\gamma,\lambda\in\square$, it can be classified according to its $r_A$ and $r_B$ value as defined above from the localisation centres of the eigenstates, we will denote such a quartet as $\square_{r_Ar_B}$.

With the organisational principle at hand, a natural step towards understanding $\pusq(\omega)$ in Eq.~\ref{eq:P-sq} is the distribution of $\tabcd$ over all quartets $\square_{r_Ar_B}$, which we define as 
\eq{
\purarb(\omega) = \frac{1}{N_{r_Ar_B}}\sum_{\substack{\alpha,\beta,\gamma,\lambda\\\in\square_{r_Ar_B}}}\expval{\delta(\tabcd-\omega)}\,,
\label{eq:pu-rarb}
}
where $N_{r_Ar_B} = 2^{2L-r_A-r_B}$ is the number of special quartets $\square_{r_Ar_B}$. This counting can be simply understood as follows. For a given $i_A^\ast$, there are $2^{L_A-r_A}$ choices of $j_A^\ast$ where the spin configuration is the same for all the $r_A$ sites from the bipartition. Similarly, for a given $i_B^\ast$, the number of states $j_B^\ast$ which has the same configuration on $r_B$ sites from the bipartition is $2^{L_B-r_B}$. As such, with the total number of choices of $i_A^\ast (i_B^\ast)$ being $2^{L_A} (2^{L_B})$, the total number of special quartets $\square_{r_Ar_B}$ is therefore $2^{L_A}2^{L_A-r_A}\times 2^{L_B}2^{L_B-r_B}$ which yields the expression above.

\begin{figure}[!b]
\includegraphics[width=\linewidth]{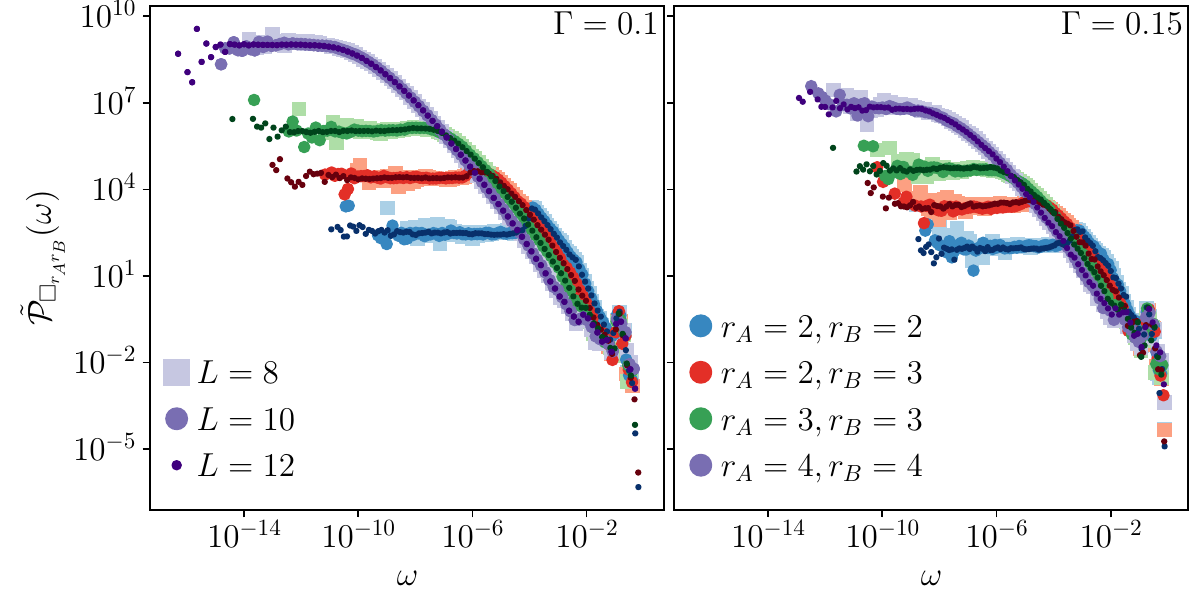}
\caption{Evidence for the fact that $\purarb$, defined in Eq.~\ref{eq:pu-rarb}, for a given $r_A$ and $r_B$ is independent of the system size $L$. The two panels correspond to the two values of $\Gamma=0.1$ and $0.15$. Different colours correspond to different values of the pair of $r_A$ and $r_B$ as indicated in the legend to the right panel whereas different markers (and intensities) correspond to different $L$ as specific in the legend in the left panel.}
\label{fig:P-theta-L-indep}
\end{figure}

Since $\tabcd$ is expected to be dominated by the {\it effective} interaction between the sites which are at a distance $r_A+r_B$ from each other (see right panel of Fig.~\ref{fig:Sq-cartoon}), we expect $\purarb(\omega)$ in Eq.~\ref{eq:pu-rarb} to be independent of $L$, as the distance itself is independent of $L$ and the microscopic interactions are strictly local. Evidence of this is provided by the results in Fig.~\ref{fig:P-theta-L-indep}. The results therein show that, for a given $r_A$ and $r_B$, the data for $\purarb$ for different $L$ indeed fall on top of each other.

\begin{figure}
\includegraphics[width=\linewidth]{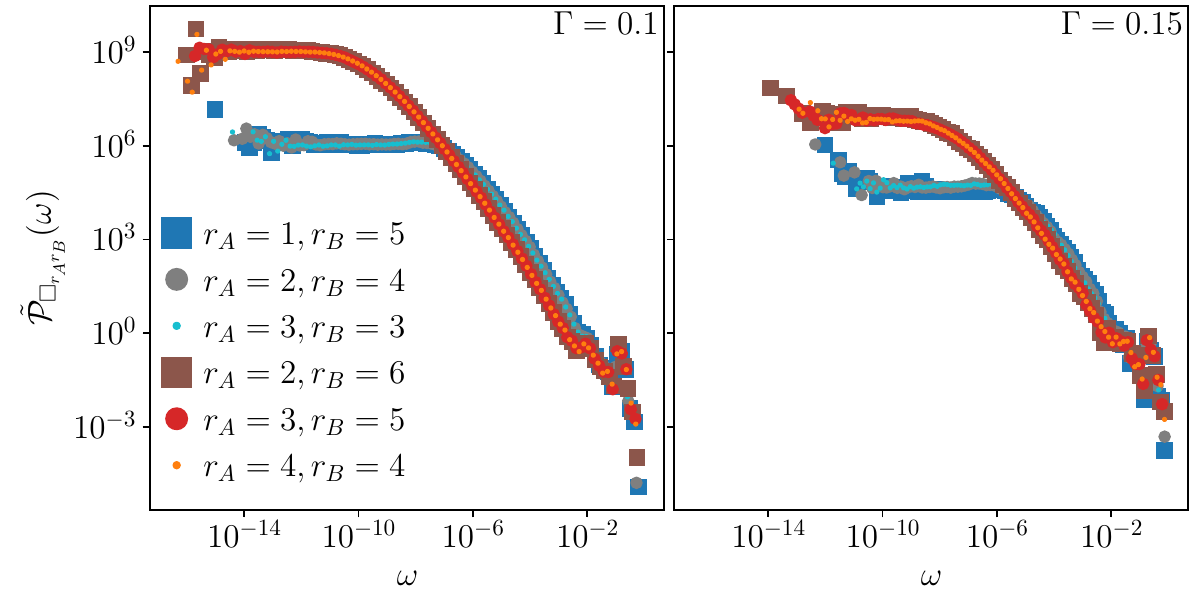}
\caption{Evidence for the fact that $\purarb$, defined in Eq.~\ref{eq:pu-rarb}, is in fact a function of $r=r_A+r_B$. Sets of data for different $r_A$ and $r_B$ which correspond to the same $r$ fall on top of each other. The two panels correspond to the two values of $\Gamma=0.1$ and $0.15$ and the shown data is for $L=12$.
}
\label{fig:P-theta-r-indep}
\end{figure}

The same argument as above also suggests that $\purarb$ for a given $r_A$ and $r_B$ should depend solely on $r\equiv r_A+r_B$. That this is indeed the case is shown by the results in Fig.~\ref{fig:P-theta-r-indep}, where it is clear that the data for $\purarb(\omega)$ are identical for different pairs of $r_A$ and $r_B$ which correspond to the same $r$. It will therefore be useful to denote by $\square_r$ the special dominant quartets $\square_{r_Ar_B}$ with $r_A+r_B=r$ and define a distribution analogous to Eq.~\ref{eq:pu-rarb} as 
\eq{
\pur(\omega) = \frac{1}{N_r}\sum_{\substack{\alpha,\beta,\gamma,\lambda\\\in\square_{r}}}\expval{\delta(\tabcd-\omega)}\,,
\label{eq:pur}
}
where $N_r$ is the number of special quartets $\square_{r}$ with
\eq{
N_r = 2^{2L-r}\times \begin{cases}
    r-1\,;&~r\leq L/2\\
    L-r+1\,;&~r>L/2
\end{cases}\,,
\label{eq:Nr-r-dep}}
where the second factor is nothing but the number of integer pairs $(r_A,r_B)$ which sum to $r$ and the first term is just the number of special quartets with a specific $r_A$ and $r_B$ which sums to $r$.
 
\begin{figure}
\includegraphics[width=\linewidth]{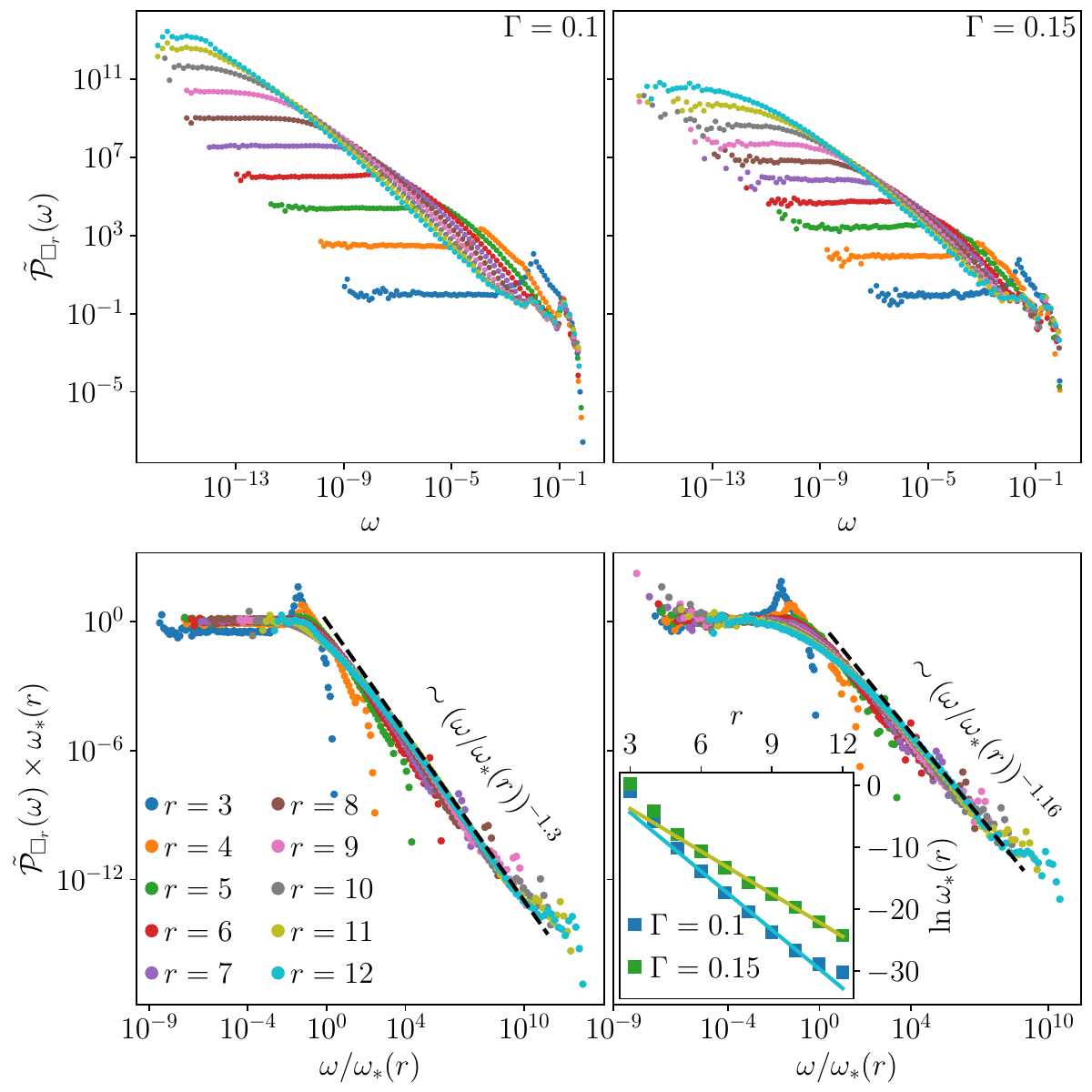}
\caption{Numerical results for $\pur(\omega)$ defined in Eq.~\ref{eq:pur}. The top panels shows the raw data for several values of $r$ as marked in the legend. The lower panels show the same data but with $\omega$ rescaled with $\omega_\ast(r)$ and the the distribution $\pur(\omega)$ rescaled with $f(r)$ where $f(r)=\omega_\ast^{-1}(r)$. The excellent scaling collapse of the data for different $r$ provides evidence for the scaling form in Eq.~\ref{eq:pu-scaling}. The left and right columns of main panels correspond to $\Gamma=0.1$ and $\Gamma=0.15$ respectively and the data is for $L=12$. The inset in the lower right panel shows that $\omega_\ast(r)$ decays exponentially with $r$ (see Eq.~\ref{eq:omega-ast}) with $\xi=0.32$ for $\Gamma=0.1$ and $\xi=0.44$ for $\Gamma=0.15$.}
\label{fig:P-theta-r-rescaled}
\end{figure}

We have already established that $\pur(\omega)$ in Eq.~\ref{eq:pur} independent of system size; however we still need to understand its dependence on $r$ and $\omega$. The data for $\pur(\omega)$ as a function of $\omega$ for several values of $r$ is shown in Fig.~\ref{fig:P-theta-r-rescaled} (top panels). There are two salient features of the data worth noting. For every $r$, there seems to exists a characteristic scale $\omega_\ast(r)$, above which $\pur(\omega)$ decays with $\omega$ as a power-law with the exponent seemingly, and crucially, independent of $r$. At the same time, below this $\omega_\ast(r)$, $\pur(\omega)$ is independent of $\omega$ and has a plateau at the value which appears to grow exponentially with $r$. This motivates a scaling form 
\eq{
\pur(\omega) = f(r)\,{\cal F}\left(\frac{\omega}{\omega_\ast(r)}\right)\,,
\label{eq:pu-scaling}
}
with the asymptotic behaviour of the scaling function,
\eq{
{\cal F}(x) = \begin{cases}
x^{-\mu}\,;&~x\gg 1\\
1\,;&~x\to 0
\end{cases}\,.
\label{eq:asymp}
}
Since $\pur(\omega)$ is a normalised distribution, it automatically mandates that $f(r) = \omega_\ast^{-1}(r)$. The data shown in the lower panels of Fig.~\ref{fig:P-theta-r-rescaled} shows that the scaling form in Eq.~\ref{eq:pu-scaling} with the scaling function in Eq.~\ref{eq:asymp} is an excellent description of $\pur(\omega)$. The power-law exponent $\mu>1$ and it decreases with increasing $\Gamma$.
The inset in the lower right panel shows that $\omega_\ast(r)$ does decay exponentially with $r$,
\eq{
\omega_\ast(r)= c\,\exp(-r/\xi)\,,
\label{eq:omega-ast}
}
where $\xi$ is a characteristic lengthscale. The data in the inset in Fig.~\ref{fig:P-theta-r-rescaled} shows that $\xi$ increases with $\Gamma$.

\subsection{Emergence of logarithmic growth of entanglement \label{sec:log-growth}}
We next show how the scaling form for $\pur$ discussed above leads to the power-law in time decay of the purity, or equivalently, the logarithmic in time growth of the second R\'enyi entropy. Recall from Eq.~\ref{eq:pu-pusq} that $\langle\puo(\omega)\rangle=\pusq(\omega)$ which can be written as a sum,
\eq{
\langle\puo(\omega)\rangle = \frac{1}{\nh^2}\sum_{r=2}^L N_r\pur(\omega)\,.
\label{eq:pu-r-sum}
}
The scaling form in Eq.~\ref{eq:pu-scaling} implies that for a given $\omega$, there exists a characteristic scale $r_\ast(\omega)$ such that 
\eq{
\pur(\omega)=
\begin{cases}
\omega^{-1}_\ast(r)\,;&~~r<r_\ast(\omega)\\
\omega^{-\mu} [\omega_\ast(r)]^{-1+\mu}\,;&~~r>r_\ast(\omega)
\end{cases}\,,
\label{eq:pur-func}
}
where
\eq{
    r_\ast(\omega) = \xi |\ln(\omega/c)|\,.
    \label{eq:r-ast}
}
Using the two cases in the equation above, the sum in the right-hand side in Eq.~\ref{eq:pu-r-sum} can be split into two as 
\eq{
\langle\puo(\omega)\rangle = \frac{1}{\nh^2}\bigg[&\sum_{r=2}^{r_\ast(\omega)}N_r\omega^{-1}_\ast(r)+\nonumber\\&\omega^{-\mu}\sum_{r=r_\ast(\omega)+1}^L N_r[\omega_\ast(r)]^{-1+\mu}\bigg]\,.
\label{eq:pu-sum-two-terms}
}
While the summations in Eq.~\ref{eq:pu-sum-two-terms} can be done exactly, as detailed in Appendix \ref{app:sum-detail}, here we only sketch the derivation and discuss the physical import of the result. 
A couple of key assertions that we make here are  $\mu>1$ and $\xi<1/\ln 2$; the numerical results in Fig.~\ref{fig:P-theta-r-rescaled} indeed show that they are satisfied. However, we will see {\it post facto} that violating these bounds leads to results which are unphysical or incompatible with locality.

In the first summation in Eq.~\ref{eq:pu-sum-two-terms}, note that the summand, $\sim 2^{-r}e^{r/\xi}$ is an exponentially increasing function of $r$ as $\xi<1/\ln 2$, and hence the sum is dominated by $r=r_\ast(\omega)$. As such, the first sum can be estimated as $2^{-r_\ast(\omega)}e^{r_\ast(\omega)/\xi}r_\ast(\omega)$ which using Eq.~\ref{eq:r-ast} is $\sim\omega^{-1+\xi\ln 2}\ln|\omega|$. On the other hand, in the second summation, by virtue of $\mu>1$ and $\xi<1/\ln 2$, the summands are an exponentially decreasing function of $r$, and hence the summation is expected to be dominated by the first term $r=r_\ast(\omega)+1$, which again leads to an estimate of the sum of as $\sim\omega^{-1+\xi\ln 2}|\ln(\omega/c)|$, which is the same as the first summation. We therefore conclude that
\eq{
    \langle \puo(\omega)\rangle \sim\omega^{-1+\xi\ln 2}|\ln(\omega/c)|\,,
    \label{eq:pu-o-ans}
}
which is the precisely the power-law in $\omega$ behaviour of $\puo(\omega)$ we sought to understand but with an additional logarithmic correction. 
Note that the result in Eq.~\ref{eq:pu-o-ans} implies that for $\puo(\omega)$ to decay with $\omega$, we necessarily require $\xi\ln 2<1$ which validates one the two assertions made earlier. Also note that, if $\mu<1-\xi\ln 2$, the second sum in Eq.~\ref{eq:pu-sum-two-terms} is dominated by $r\sim L$ at finite $\omega$, which is clearly incompatible with dynamics induced by a local unitary time-evolution implying $\mu$ must satisfy $\mu>1-\xi\ln 2$. Since the theory must remain valid in the deep MBL phase (and in fact, get better on going deeper) the inequality must hold as $\xi\to 0$ implying $\mu>1$ and thus validating the second assertion.

In the time domain, the Fourier transform of  Eq.~\ref{eq:pu-o-ans} leads to
\eq{
\langle \pu(t)\rangle\sim t^{-\xi \ln 2}\ln (ct)\,,
\label{eq:Pt-ans}
}
which again is nothing but the power-law decay in time of the purity (Eq.~\ref{eq:Pt-power-law}) with a logarithmic correction, as seen numerically in Sec.~\ref{sec:model}. Note that $\langle S_2^{AB}(t)\rangle\approx -\log \langle \pu(t)\rangle$ (see Fig.~\ref{fig:Pt-S2t}), we conclude that 
\eq{
\langle S_2^{AB}(t)\rangle \sim (\xi \ln 2) \ln t  - \ln \ln (ct) \overset{t\gg 1}{\sim} (\xi \ln 2) \ln t\,,
}
which is the logarithmic growth of the entanglement entropy characteristic of the MBL regime. From the data in Fig.~\ref{fig:P-theta-r-rescaled}, we have  $\xi=0.32$ for $\Gamma=0.1$ and $\xi=0.44$ for $\Gamma=0.15$. For these values, the power-law-exponent, $\xi\ln 2$ in Eq.~\ref{eq:Pt-ans}, takes on values approximately 0.22 and 0.31 for $\Gamma=0.1$ and $0.15$ respectively. The numerical results in Fig.~\ref{fig:Pt-S2t} on the other hand estimate the exponent $a$ to be 0.17 and 0.24. We note that while the two estimates are reasonably close, $a \lessapprox \xi \ln 2$ which we attribute to the logarithmic correction in Eq.~\ref{eq:Pt-ans}.
 In particular, taking into account the logarithmic correction while analysing the data in Fig.~\ref{fig:Pt-S2t} does yield power-law exponents in excellent agreement with the theoretical prediction, see Appendix~\ref{app:logcorr} for details.

\subsection{Spacetime picture of entanglement growth \label{sec:spcaetime}}

An important physical import of the above analysis is that it provides a spacetime picture of the entanglement growth as follows. For a given $\omega$, the sum in Eq.~\ref{eq:pu-r-sum} is dominated by a single $r = r_\ast(\omega)$. In other words, the dynamics of the purity at frequency $\omega$ is encoded in the spectral correlations of the form in Eq.~\ref{eq:P-sq} within the special quartets $\square_{r_\ast(\omega)}$ with $r_\ast(\omega)$ given by Eq.~\ref{eq:r-ast}. This suggests that that at any time $t$, there is a characteristic lengthscale $d(t)\equiv r_\ast(\omega=2\pi/t)\sim \xi\ln t$  which contributes dominantly to the growth of entanglement at time $t$. Since degrees of freedom further away from the bipartition get entangled at later times, loosely speaking, $d(t)$ can be understood as a lengthscale degrees of freedom within which are entangled at time $t$.
As the total entanglement carried in such a case will be proportional to $d(t)$, this presents an alternative perspective to infer the logarithmic growth of entanglement entropy. In addition, this could also be understood as the manifestation of the logarithmic light-cone of operator spreading in the MBL regime~\cite{chen2016universal,chen2017out,huang2017out,deng2017logarithmic}.

\begin{figure}
\includegraphics[width=\linewidth]{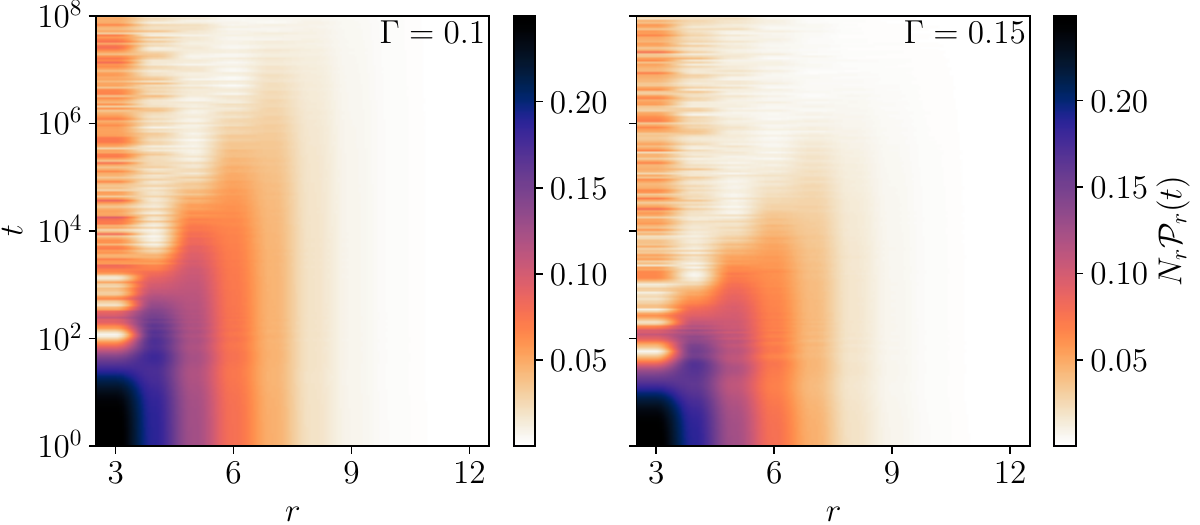}
\caption{The spacetime picture of entanglement growth obtained numerically. The heatmap shows $N_r{\cal P}_{r}(t)$ (defined in Eq.~\ref{eq:prt-sum}) in the $(r,t)$ plane. Note that the $t$ axis is logarithmic. As such the linear front seen in the plots corresponds to a logarithmic front in time. Data is for $L=12$ and the left and right panels correspond to $\Gamma=0.1$ and $0.15$ respectively. Also note that the wavefront is slower for smaller $\Gamma$ as one expects deeper in the MBL phase.}
\label{fig:Nr-Prt-numerical}
\end{figure}

Understanding the dynamics of purity resolved in real-space distance leads to further insights. The purity in time, using Eq.~\ref{eq:pu-r-sum}, can be expressed as 
\eq{
\langle\pu(t)\rangle = \frac{1}{\nh^2}\sum_{r=2}^L N_r {\cal P}_r(t)\,,
\label{eq:prt-sum}
}
where ${\cal P}_r(t)$ is the Fourier transform of $\pur(\omega)$.
 We first present microscopic results for $N_r{\cal P}_r(t)$  as a heatmap in $r$ and $t$, obtained directly from their numerical computation. The results are shown in Fig.~\ref{fig:Nr-Prt-numerical} where it is clear that there is a `feature' which spreads in space logarithmically in time (note that the time axes in the plots in Fig.~\ref{fig:Nr-Prt-numerical} are logarithmic). In particular, it is clear that at a given $t$ there is a peak in $N_r{\cal P}_r(t)$ at a specific value of $r$ which grows logarithmic with $t$; this is nothing but the aforementioned $d(t)$. This provides us with the notion of an `entanglement front'.

\begin{figure}
\includegraphics[width=\linewidth]{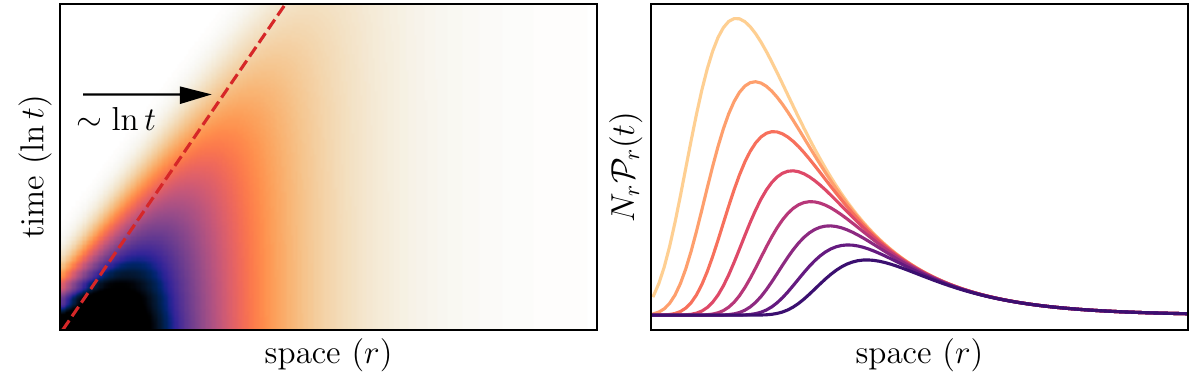}
\caption{The spacetime picture of entanglement growth obtained from our theory. The left panel shows a heatmap in spacetime of $N_r{\cal P}_r(t)$, defined in Eq.~\ref{eq:prt-sum}, where we used the expression in Eq.~\ref{eq:Nr-r-dep} for $N_r$ and that in Eq.~\ref{eq:str-exps} for ${\cal P}_{r}(t)$. The wavefront spreads logarithmically in time as indicated by the red dashed line. The right panel shows wavefronts in space at fixed times (darker colours show later times) which are horizontal cuts of the figure in the left panel.}
\label{fig:spacetime}
\end{figure}

To understand analytically the origins of this, note that the functional form of $\pur(\omega)$ in Eq.~\ref{eq:pur-func} suggests that ${\cal P}_r(t)$ decays like a stretched-exponential in time~\cite{wuttke2012laplace},
\eq{
{\cal P}_r(t)\sim \exp\left[-\left(\frac{t}{t_\ast(r)}\right)^z\right]\,;\quad t_\ast(r)\sim \omega_\ast(r)^{-1}\,.
\label{eq:str-exps}
}
 An analytic expression of the kind above along with $N_r$ given by Eq.~\ref{eq:Nr-r-dep} allows us to show the entanglement wavefront in a much cleaner fashion for arbitrarily large systems, simply by plotting the closed form expressions. This is done in Fig.~\ref{fig:spacetime}.
Since $t_\ast(r)$ grows exponentially with $r$, the temporal dependence of ${\cal P}_r(t)$ naturally slows down with increasing $r$. At the same time, the number of quartets with $r$ also decays exponentially with $r$, Eq.~\ref{eq:Nr-r-dep}. The power-law decay of $\langle\pu(t)\rangle$ emerges from the collective effect of the stretched-exponential decays in Eq.~\ref{eq:str-exps} with hierarchically growing timescales, $t_\ast(r)\sim e^{r/\xi}$ and the hierarchically decreasing $N_r\sim e^{-r\ln 2}$. This arises simply from the fact that the sum in Eq.~\ref{eq:prt-sum} is dominated by a `saddle-point', $r_\ast(t)\sim \xi \ln t$.  It is also interesting to note that the phenomenology is quite similar to that of hierarchically constrained dynamics for glassy relaxation~\cite{palmer1984models}. 

From the space-resolved picture discussed above, the entanglement wavefront can be understood as follows. The partial sum of Eq.~\ref{eq:prt-sum}, $\sum_{r=2}^R N_r{\cal P}_r(t)$ can be understood to contain the information of the decay of bipartite purity between $A$ and $B$ carried by degrees of freedom within a distance $2R$ from the bipartition. $N_r{\cal P}_r(t)$ is therefore the contribution to the decay in purity at time $t$ at distance $r$ from boundary. However, the aforementioned $r_\ast(t)$ suggests that there is a well-defined wavefront of entanglement spreading,{as seen numerically in Fig.~\ref{fig:Nr-Prt-numerical}} and theoretically in Fig.~\ref{fig:spacetime}. For a given $t$, the wavefront decays exponentially with $r$ away from $r_\ast(t)$ for $r>r_\ast(t)$ due to the exponential decay with $r$ of $N_r$. On the other hand, the hole in the profile for $r$ away from $r_\ast(t)$ with $r<r_\ast(t)$ is purely due to the fact that degrees of freedom within distance $r$ of the bipartition are strongly entangled and the effective contribution to the purity from them has decayed to extremely small values.

\subsection{$\ell$-bits redux \label{sec:ellbit}}

Having understood the microscopic origins of the logarithmic growth of entanglement in the MBL regime, it is useful to discuss how the microscopic theory is connected to the phenomenological $\ell$-bit picture, and in fact goes beyond it. The $\ell$-bit picture suggests that the effective Hamiltonian encoding the interactions between the local integrals of motion (denoted by $\tau^z_i$) is of the form
\eq{
H_{\ell\text{-bit}}=\sum_i h_i\tau^z_i + &\sum_{j>i}J_{ij}e^{-\frac{j-i}{\zeta_{ij}}}\tau^z_i\tau^z_j + \nonumber\\&\sum_{k>j>i}J_{ijk}e^{-\frac{k-i}{\zeta_{ijk}}}\tau^z_i\tau^z_j\tau^z_k\cdots\,.
\label{eq:ham-lbit}
}
A simplifying assumption commonly made within the $\ell$-bit picture is that the lengthscale $\zeta$ can be taken to be a characteristic localisation length of the MBL system which depends only on the disorder strength. The energyscale which therefore controls the entanglement spreading until distance $r$ from the bipartition is $\sim e^{-r/\zeta}$ and the corresponding timescale is $\sim e^{r/\zeta}$. In other words, at time $t$, the `entanglement wavefront' reaches a distance $d(t) \sim \zeta \ln t$. Since the entanglement is proportional to this volume, $S(t)\sim d(t)$ one obtains the logarithmic growth of entanglement $S(t)\sim \zeta \ln t$~\cite{serbyn2013universal}.
While such a picture recovers the logarithmic growth of entanglement in time, it is insufficient to describe the distributions of relevant energy- and timescales or the spacetime profile of entanglement growth. This picture also fails to take into account the several corrections to the energyscales arising from the interactions between more distant degrees of freedom~\cite{serbyn2013universal}.
This is fundamentally due to the absence of an understanding of the distribution of the lengthscales $\{\zeta\}$ which in turn is rooted in an absence of an explicit construction of the $\ell$-bit Hamiltonian starting from a microscopic theory.

This is where our theory goes beyond the conventional $\ell$-bit picture. Note that the energyscale that controls the entanglement spreading at distance $r$ from the bipartition is the $\tabcd$ for $\alpha,\beta,\gamma,\lambda\in\square_r$. Crucially, in our approach, the set of $\{\tabcd\}$ and their distributions can be computed explicitly with the latter given simply by $\pur(\omega)$ defined in Eq.~\ref{eq:pur}. Since this is computed exactly, for a given $r$, it manifestly contains not just the leading contribution due to the interactions between degrees of freedom separated by $r$ but also the smaller corrections due to the interactions between degrees of freedom further away. 

In fact, since the spacetime picture in Fig.~\ref{fig:spacetime} follows directly from the distribution $\pur(\omega)$, the logarithmic spreading of the entanglement wavefront and the stretched-exponential decay of ${\cal P}_r(t)$ can be attributed to the collective effect of the plethora of energyscales and timescales that emerge from the all the interactions between degrees of freedom at different distances. In this way, the microscopic picture developed in this work presents a much richer picture of the entanglement growth in an MBL phase than what is possible within a $\ell$-bit picture.

An alternative viewpoint could be that the interactions between the faraway degrees of freedom are irrelevant and the distribution $\pur(\omega)$ originates purely from a distribution of the lengthscales $\{\zeta\}$. However, positing that $\tabcd \sim e^{-r/\zeta}$ for $\alpha,\beta,\gamma,\lambda\in\square_r$, the distribution $\pur$ can be transformed to yield a distribution $P_{\zeta}$ for $\zeta$ -- assuming the form in Eq.~\ref{eq:pur-func} the latter exhibits a power-law tail $P_{\zeta}\sim \zeta^{-2}$. The tail in the distribution rules out the presence of finite, mean localisation length.  However, more importantly, this does not preclude the logarithmically slow in time growth of entanglement and hence a characteristic of the MBL regime.

\section{Infinite-time purity \label{sec:inftime}}
While the focus of this work has been on the dynamics of purity, as a matter of completeness, we discuss how the eigenstate correlations also encode the infinite-time saturation value of the purity. Specifically, we will discuss the system-size scaling of $\pusat$ in Eq.~\ref{eq:pu-sat}.

For an {\it interacting} quantum many-body system of finite-size without any conservation laws, there are no degeneracies in the eigenvalues. Moreover, the probability that $\tabcd=0 [{\rm mod} 2\pi]$ where all four of $\al$, $\be$, $\ga$, and $\la$ are different is measure zero. Hence the only distinct possibilities for having $\tabcd=0$ are $\alpha=\beta=\ga=\la$, $\alpha=\beta$ and $\ga=\la$ with $\al\neq \ga$, and $\alpha=\ga$ and $\be=\la$ with $\al\neq \be$. These conditions can be used in Eq.~\ref{eq:pu-sat} to give
\eq{
    \begin{split}
    \pusat = \frac{1}{\nh^2}\sum_{\al,\be}&[(V_{\al\be\al\be}+V_{\al\al\be\be})\times\\&(1+V_{\al\be\al\be}+V_{\al\al\be\be})]-\frac{2}{\nh^2}\sum_{\al}V_{\al\al\al\al}^2\,.
    \end{split}
    \label{eq:pu-sat-V}
}
The terms on the right-hand side of Eq.~\ref{eq:pu-sat-V} satisfy certain sum rules and inequalities.
From the orthonormality of the eigenstates, $\braket{\alpha}{\beta}=\delta_{\alpha\beta}$, it can be straightforwardly shown that
\eq{
\sum_{\alpha,\beta}V_{\alpha\alpha\beta\beta}  = \sum_{\alpha,\beta}V_{\alpha\beta\alpha\beta} = \nh^{3/2}\,.
\label{eq:Vabab-sumrule}
}
Also note that $0<V_{\alpha\beta\alpha\beta},V_{\al\al\be\be}<1$. The first inequality is trivially manifest as $\Vabab= \sum_{i_B,j_B}|\sum_{i_A}\alpha_{i_Ai_B}\beta^\ast_{i_Aj_B}|^2$ and similarly for $V_{\al\al\be\be}$. At the same time since $V_{\alpha\alpha\beta\beta},V_{\alpha\alpha\beta\beta}$ is sum over products of normalised eigenstate amplitudes, the second inequality also follows. This naturally implies that 
\eq{
\sum_{\alpha,\beta}|V_{\alpha\alpha\beta\beta}|^2\leq \sum_{\alpha,\beta} V_{\alpha\alpha\beta\beta}\Rightarrow \sum_{\alpha,\beta}|V_{\alpha\alpha\beta\beta}|^2\leq \nh^{3/2}\,,
\label{eq:V-ineq}
}
and similarly for $\sum_{\alpha,\beta}|V_{\alpha\beta\alpha\beta}|^2$. The same argument goes through for the quantity $\sum_{\alpha,\beta}V_{\al\be\al\be}V_{\al\al\be\be}$ implying $\sum_{\alpha,\beta}V_{\al\be\al\be}V_{\al\al\be\be}< \nh^{3/2}$. Finally, note that $V_{\al\al\al\al}$ is nothing but the purity $\pu$ of the eigenstate $\ket{\alpha}$. The area-law entanglement of MBL eigenstates then means $V_{\al\al\al\al}\sim O(1)$ which in turn implies
\eq{
\sum_\alpha V_{\al\al\al\al}^2 \sim O(\nh)\,.
\label{eq:eig-pur}
}
The sum rules and inequalities in Eq.~\ref{eq:Vabab-sumrule}-\ref{eq:eig-pur}, therefore suggest that it is the first term in Eq.~\ref{eq:pu-sat-V}, $\pusat\approx\nh^{-2}\sum_{\alpha,\beta}(V_{\al\be\al\be}+V_{\al\al\be\be})$ which dominates the scaling of $\pusat$ with system size. As such, we have
\eq{
\pusat\overset{L\gg 1}{\approx} 2\nh^{-1/2}\Rightarrow S_2^{AB}(t\to\infty)\approx \frac{L}{2}\ln 2\,.
}
Note that the entanglement entropy saturates to the maximal (thermal at infinite temperature in this case) volume law in this case. This is due to the fact that initial states of the form in Eq.~\ref{eq:psi0-product} are fully ergodic in the eigenbasis of $U_F$. More commonly considered initial states, such as $\sigma^z$-product states (or computational basis states) are typically multifractal in the basis of MBL eigenstates which leads to the sub-thermal volume law saturation of the entanglement entropy~\cite{bardarson2012unbounded,serbyn2013universal}.

\subsection*{Digression: area-law saturation of entanglement in Anderson localised systems}

It is important to note here that Eq.~\ref{eq:pu-sat-V} is {\it not} valid for a non-interacting Anderson localised system. As we discuss next, it is this issue that lies at the heart of the entanglement entropy saturating to an area-law value in an Anderson localised system compared to volume-law saturation in an MBL system. For an Anderson localised system, let us denote the single-particle eigenstates as $\{\ket{a}\}$ and the corresponding eigenenergies as $\{\varepsilon_a\}$. The many-body eigenvalues are then simply given by
\eq{
    \theta_\alpha = \sum_{a=1}^L \eta_\alpha^a\varepsilon_a\,,
}
where $\eta_\alpha^a=1,0$ is the occupancy of the orbital $a$ in the eigenstate $\ket{\alpha}$. In such a case, 
\eq{
    \tabcd = \sum_{a=1}^L (\eta_\alpha^a-\eta_\beta^a-\eta_\gamma^a+\eta_\lambda^a)\varepsilon_a\,.
}
This allows for multitude of more possibilities for $\tabcd=0$ compared to the interacting MBL case. In particular, for a quartet of eigenstates $(\alpha,\beta,\gamma,\lambda)$, all of them different, if the orbitals whose occupancy are different between $\ket{\alpha}$ and $\ket{\beta}$ are the same orbitals whose occupancy are different (with the same sign) between $\ket{\gamma}$ and $\ket{\lambda}$, it leads to $\tabcd=0$, and in principle they can all contribute to the second term in $\pusat$ in Eq.~\ref{eq:pu-sat}. However, the crucial point is that the single-particle orbitals, $\{\ket{a}\}$ are all exponentially localised in space. Hence, for a quartet, if the aforementioned orbitals which differ in occupancy between $\ket{\alpha}$ and $\ket{\beta}$, and between $\ket{\gamma}$ and $\ket{\lambda}$, are sufficiently far away (further than a few single-particle localisation lengths) from from boundary between $A$ and $B$, it leads to $\Vabcd\sim 0$. On the other hand, if all of these single-particle orbitals straddle the boundary between $A$ and $B$, it leads to $\Vabcd \sim O(1)$. It is straightforward to see that the number of such quartets is $\propto \nh^2$ with a the proportionality factor being the single-particle localisation length. Using this in the second term in Eq.~\ref{eq:pu-sat}  leads to an infinite-time purity which is proportional to the single-particle localisation length, and hence leads to an area-law saturation of the entanglement entropy.

It is interesting to note that while the many-body eigenstates individually are qualitatively similar in many way in both the interacting and non-interacting cases, it is the stark difference in the spectral correlations that leads to the volume-law saturation of entanglement entropy in the interacting case, and area-law in the non-interacting case.

\section{Summary and Outlook \label{sec:conclusion}}

Let us summarise the main results of the paper briefly. We developed a concrete relation between the dynamics of bipartite entanglement entropy, of states as well as that of the time-evolution operator, and the minimal eigenstate- and spectral correlations that encode them in a generic, interacting quantum system. In particular, we focussed on the second R\'enyi entropy of entanglement or equivalently the subsystem purity. The eigenstate correlations involve quartets of at least four eigenvectors and hence, manifestly go beyond those mandated by the ETH or lack thereof in strongly disordered, ergodicity-broken systems. Using this lens of eigenstate and spectral correlations, we constructed a microscopic theory of the logarithmic growth in time of entanglement entropy in the MBL regime at strong disorder. One key aspect of the theory was the identification that for strongly disordered systems, the dynamics of purity is contributed to by the eigenstate and spectral correlations of a vanishing fraction of the quartets. In particular, out of the $O(\nh^4)$ possible quartets, special set of $O(\nh^2)$ quartets dominate the dynamics of the purity. In fact, we discover that the spectral correlations within these special dominant quartets is sufficient to capture the dynamics. We uncover the defining structure of these quartets, both in real and Hilbert space, which provides us with an organising principle for these special quartets, in terms of a characteristic lengthscale. Using this, we find that there exists a hierarchy of characteristic frequency- and timescales for these quartets based on their characteristic lengthscales. The interplay of these hierarchical timescales and the number of such special quartets with a particular lengthscale eventually leads to the power-law decay in time of purity or equivalently, the logarithmic in time growth of entanglement entropy. We discussed briefly implications of our results for the phenomenological $\ell$-bit picture and also the infinite-time saturation of the purity, in particular its scaling with system size, as manifested in the eigenstate correlations.

All the results summarised above are rooted in an understanding of the eigenstate and spectral correlations involving four different eigenstates. Questions to follow up in the future therefore arise where such correlations or their generalisations appear naturally. While we focused here on the entanglement entropy of a system with no conservation laws, a similar theory can be constructed for the unresolved issue of the dynamics of number entropy in MBL systems with a conserved charge~\cite{kiefer2020evidence,luitz2020absence,ghosh2022resonance,chavez2023ultraslow}.
In order to better understand the spatiotemporal structure of statistics of eigenstates, analogous to Ref.~\cite{hahn2023statistical}, one can study the correlation $\Vabcd(\Vabcd^\prime)^\ast$ resolved in $\tabcd$ where $\Vabcd^\prime$ is the same eigenstate correlation as defined in Eq.~\ref{eq:V-abcd} but with a different bipartition of the system. In a similar spirit, one can introduce matrix elements of local operators and recast the OTOC or temporal fluctuations of local observables~\cite{serbyn2014quantum} from the point-of-view of eigenstate correlations. In the same direction of spatiotemporal structure of entanglement dynamics, it will be important to understand the origins of the putative stretched-exponential decay in time of purity resolved in $r$ (see Eq.~\ref{eq:str-exps}). Usually such anomalous decays appear due to the presence of a broad distribution of timescales due to heterogeneity; whether this is connected to the dynamical heterogeneity in the entanglement dynamics, akin to classical glasses~\cite{artiaco2022spatiotemporal}, remains a work for the future. Relatedly, the spacetime picture also raises the question that whether there exists a surface growth model for entanglement growth~\cite{nahum2018dynamics} which describes the dynamics of entanglement in the MBL phase.

\begin{acknowledgements}
We thank F. Alet, C. Artiaco, J. H. Bardarson, D. A. Ch\'avez, N. Laflorencie and I. M. Khaymovich for several useful discussions. SR acknowledges support from SERB-DST, Government of India under Grant No. SRG/2023/000858, from the Department of Atomic Energy, Government of India, under Project No. RTI4001, and from an ICTS-Simons Early Career Faculty Fellowship via a grant from the Simons Foundation (Grant No. 677895, R.G.).
\end{acknowledgements}

\appendix

\section{Details of averaging over initial states \label{app:psi0-avg}}
In this appendix we present the details of the averaging over initial states and the derivation of Eq.~\ref{eq:I-abcd-avg}. The initial states we considered are random product states between $A$ and $B$ as defined in Eq. \ref{eq:psi0-product}. where, $\ket{\psi_0^A}$ and $\ket{\psi_0^B}$ are defined as follows,
\eq{\begin{split}
\ket{\psi_0^A}&=\sum_{i_A}\phi_{i_A}\ket{i_A}\,,\\
\ket{\psi_0^B}&=\sum_{i_B}\phi_{i_B}\ket{i_B}\,,
\end{split}}
where $\phi_{i_A} (\phi_{i_B})$ are independent complex Gaussian random numbers with zero mean and standard deviation $N_{{\cal H}_A}^{-1/2} (N_{{\cal H}_B}^{-1/2})$ so as to preserve normalisation of the state. Formally,
\eq{
\expval{\phi_{i_A}}=0\,;\quad\expval{\phi_{i_A}\phi_{j_A}^\ast}=N_{{\cal H}_A}^{-1}\delta_{i_Aj_A}\,.
}
Since the random states are Gaussian it naturally implies
\eq{
\mathbb{E}[\phi_{i_A}\phi_{j_A}^*\phi_{k_A}^*\phi_{l_A}]_{\{\phi_{i_A}\}}&=\frac{1}{N_{{\cal H}_A}^2}[\delta_{i_Aj_A}\delta_{k_Al_A}+\delta_{i_Ak_A}\delta_{j_Al_A}]\,,
\label{eq:phi-avg}
}
and similarly for $B$.

Now, $\mathcal{I}^{\al\be\ga\la}_{\psi_0}$ from Eq.~\ref{eq:I-abcd-psi0} can be written as 
\begin{widetext}
\eq{\mathcal{I}^{\al\be\ga\la}_{\psi_0}=\sum_{\substack{i_A,j_A,k_A,l_A, \\ 
i_B,j_B,k_B,l_B}}&\alpha_{i_Ai_B}^*\beta_{j_Aj_B}\ga_{k_Ak_B}\la_{l_Al_B}^*\times(\phi_{i_A}\phi_{j_A}^*\phi_{k_A}^*\phi_{l_A})\times(\phi_{i_B}\phi_{j_B}^*\phi_{k_B}^*\phi_{l_B})\,,\label{I-abcd-phiA-phiB}
}
Since $\ket{\psi_0^A}$ and $\ket{\psi_0^B}$ are independent of each other, the averages over $\{\phi_{i_A}\}$ and $\{\phi_{i_B}\}$ can be done separately, such that
\eq{\mathbb{E}[\mathcal{I}^{\al\be\ga\la}_{\psi_0}]_{\psi_0}&=\sum_{\substack{i_A,j_A,k_A,l_A, \\ 
i_B,j_B,k_B,l_B}}\alpha_{i_Ai_B}^*\beta_{j_Aj_B}\ga_{k_Ak_B}\la_{l_Al_B}^*\times\mathbb{E}[\phi_{i_A}\phi_{j_A}^*\phi_{k_A}^*\phi_{l_A}]_{\{\phi_{i_A}\}}\times\mathbb{E}[\phi_{i_B}\phi_{j_B}^*\phi_{k_B}^*\phi_{l_B}]_{\{\phi_{i_B}\}}\\
&=\frac{1}{\nh^2}\sum_{\substack{i_A,j_A,k_A,l_A, \\ 
i_B,j_B,k_B,l_B}}\alpha_{i_Ai_B}^*\beta_{j_Aj_B}\ga_{k_Ak_B}\la_{l_Al_B}^*\times[\delta_{i_Aj_A}\delta_{k_Al_A}+\delta_{i_Ak_A}\delta_{j_Al_A}]\times[\delta_{i_Bj_B}\delta_{k_Bl_B}+\delta_{i_Bk_B}\delta_{j_Bl_B}]
\label{eq:I-abcd-phiA-phiB-avg}\,,
}
where in the second line we used Eq.~\ref{eq:phi-avg}.
\end{widetext}
The combinations of the Kronecker-delta functions in Eq.~\ref{eq:I-abcd-phiA-phiB-avg} finally leads to 
\begin{align}
   \mathbb{E}[\mathcal{I}^{\al\be\ga\la}_{\psi_0}]_{\psi_0}=\frac{1}{\nh^2}[\delta_{\al\be}\delta_{\ga\la}+\delta_{\al\ga}\delta_{\be\la}+V_{\al\be\ga\la}^*+V_{\al\ga\be\la}^*]\,,
\end{align}
which is the content of Eq.~\ref{eq:I-abcd-avg}.

\section{Evidence for expression for $\pudyn$ \label{app:pudyn}}
\begin{figure}[!t]
    \centering
    \includegraphics[width=\linewidth]{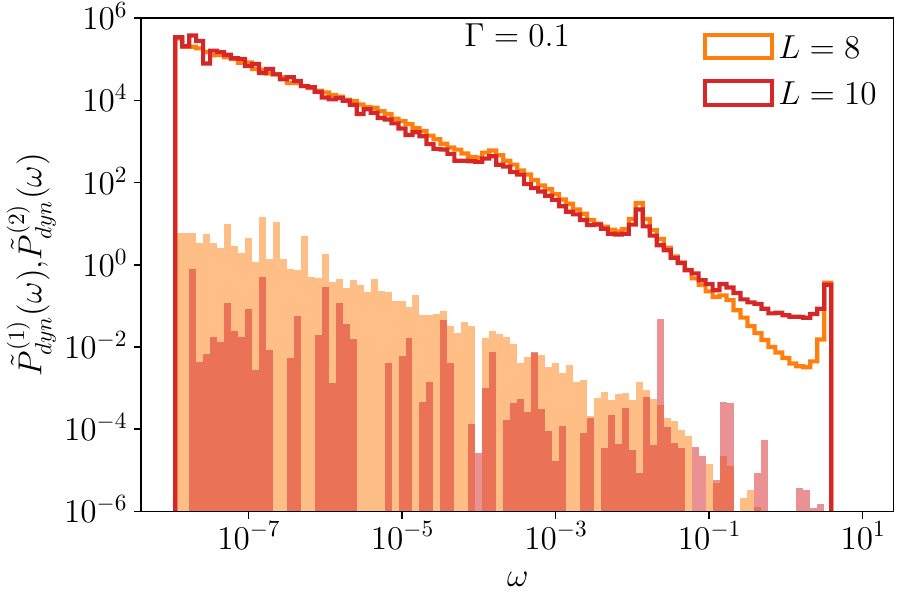}
    \caption{Comparison of $\pudyn^{(1)}(\omega)$ (lines) and $\pudyn^{(2)}(\omega)$ (filled) defined in Eq.~\ref{eq:pdyn1} and Eq.~\ref{eq:pdyn2} which shows that $\pudyn^{(1)}(\omega)\gg \pudyn^{(2)}(\omega)$ for all $\omega$ and that the inequality gets stronger with increasing $L$. }
    \label{fig:P-dyn-1-2}
\end{figure}

The expression for $\pudyn$ in Eq.~\ref{eq:pu-dyn} can be split into two terms as
\eq{
\pudyn(\omega) = \pudyn^{(1)}+(\omega)\pudyn^{(2)}(\omega)\,,
}
where 
\eq{
\pudyn^{(1)}(\omega)  &\equiv \frac{1}{\nh^2}\sum_{\substack{\al,\be,\ga,\la:\\\tabcd\neq0}}\delta(\ta_{\al\be\ga\la}-\omega)|V_{\al\be\ga\la}|^2\,.\label{eq:pdyn1}\\
\pudyn^{(2)}(\omega) &\equiv \frac{1}{\nh^2}\sum_{\substack{\al,\be,\ga,\la:\\\tabcd\neq0}}\delta(\ta_{\al\be\ga\la}-\omega)V_{\al\be\ga\la}V_{\al\ga\be\la}^\ast\,,\label{eq:pdyn2}
}
and we had neglected $\pudyn^{(2)}(\omega)$.
In the main text, we provided a justification for it on general grounds based on its integral over all $\omega$ being suppressed in Hilbert space dimension. For completeness, however we show numerical evidence in support of the fact that $\pudyn^{(2)}(\omega)\ll \pudyn^{(1)}(\omega)$ for all $\omega$. In Fig.~\ref{fig:P-dyn-1-2} we show the results for $\pudyn^{(1)}(\omega)$ and $\pudyn^{(2)}(\omega)$ as a function of $\omega$ where it is clear that that latter is several orders of magnitudes smaller than the former for all $\omega$.

\section{Existence and structure of special quartets for time-independent MBL Hamiltonians \label{app:ham}}
As discussed in Sec. \ref{sec:anatomy}, the central ingredient of our theory the existence dominant quartets whose multiplicity scales as $\nh^2$ and for which $|\Vabcd|^{2}\sim O(1)$. 
While the results centred on the Floquet system defined in Eq.~\ref{eq:UF}, here we show that they hold also for a time-independent MBL Hamiltonian, thereby providing evidence that the existence and structure of the dominant quartets is genetic to MBL phases.

\begin{figure}
    \centering
    \includegraphics[width=\linewidth]{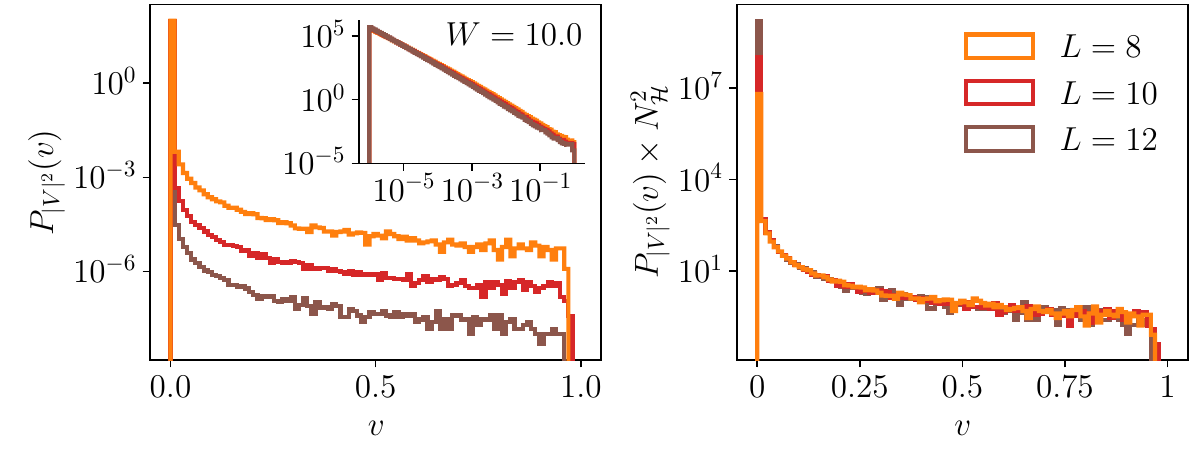}
    \caption{Data analogous to Fig.~\ref{fig:Vdist} for the time-independent Hamiltonian in Eq.~\ref{eq:ham}, again showing that there exist dominant quartets with $|\Vabcd|^2\sim O(1)$ whose multiplicity scales as $\nh^2$.}
    \label{fig:ham-dist}
\end{figure}

\begin{figure}
    \centering
    \includegraphics[width=\linewidth]{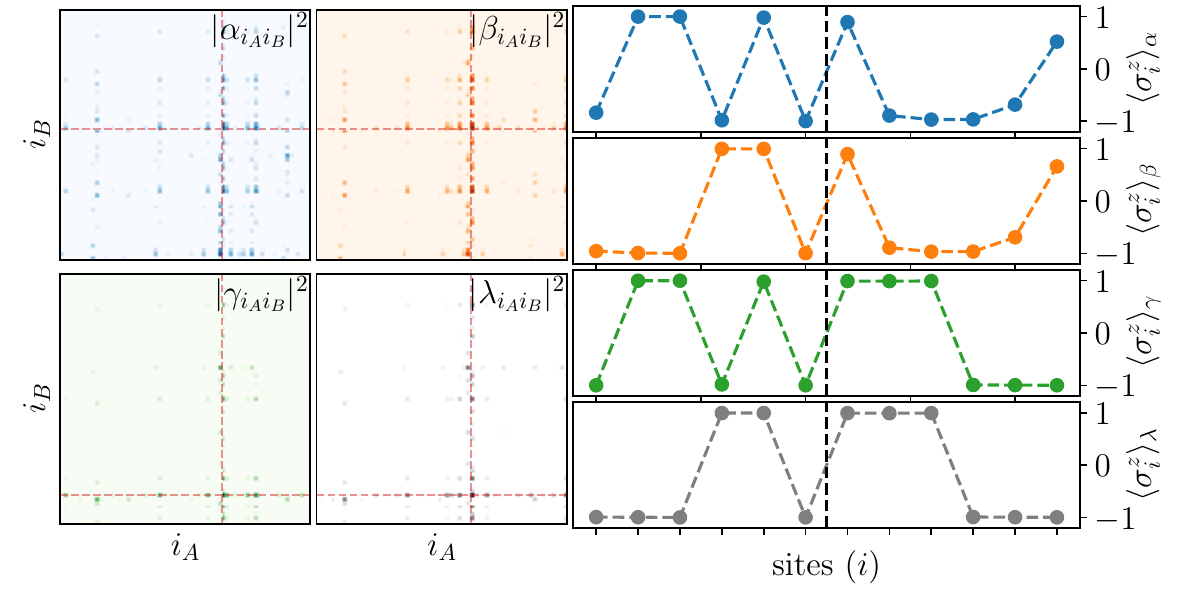}
    \caption{Data analogous to that of Fig.~\ref{fig:Sq-struture}, but for the Hamiltonian in Eq.~\ref{eq:ham}, again shows that the structure of the dominants quartets in both, Fock space and real space is the same.}
    \label{fig:ham-struc}
\end{figure}

We consider the Hamiltonian~\cite{imbrie2016many}, 
\eq{
H = \sum_{i}[J_i\sigma^z_i\sigma^z_{i+1} + h_i\sigma^z_i + \Gamma\sigma^x_i]\,,
\label{eq:ham}
}
where $J_i\in [0.8,1.2]$, $\Gamma=1$, and $h_i\in[-W,W]$ . We choose $W=10.0$ to be sufficiently deep in the MBL phase and present plots analogous to those in Fig.~\ref{fig:Vdist} and Fig.~\ref{fig:Sq-struture}. The results for the Hamiltonian \eqref{eq:ham} are shown in Fig.~\ref{fig:ham-dist} and Fig.~\ref{fig:ham-struc}. The former is identical to Fig.~\ref{fig:Vdist} whereas the latter is identical to Fig.~\ref{fig:Sq-struture}, thus showing that the multiplicity and the structure of the dominants quartets for the Hamiltonian \eqref{eq:ham} are identical to that of the Floquet system \eqref{eq:UF}.

\section{Derivation of the $\omega$-dependence of $\puo(\omega)$ \label{app:sum-detail}}
In this appendix, we provide the details of the summation in Eq.~\ref{eq:pu-sum-two-terms} and how it leads to Eq.~\ref{eq:pu-o-ans}. Note that in Eq.~\ref{eq:pu-sum-two-terms} there are two summations, which we denote as $\Sigma_1$ and $\Sigma_2$. Let us discuss the two of them separately starting with the former.

Note that for any $\omega$, that is independent of $L$, and so is $r_\ast(\omega)$ and hence $r_\ast(\omega)\ll L/2$. So for all terms in $\Sigma_1$, $N_r = 2^{2L-r}(r-1)$. We therefore have
\eq{
\Sigma_1 = \sum_{r=2}^{r_\ast(\omega)}c\,(r-1) \,e^{-r(\ln{2}-1/\xi)}\,,
}
where the summation can be performed exactly to yield
\eq{\begin{split}
    \Sigma_1=\frac{c\,e^{2/\xi}}{(e^{1/\xi}-2)^{2}}&-\frac{c^{-\xi\ln{2}}e^{2/\xi}}{(e^{1/\xi}-2)^{2}}\omega^{-1+\xi\ln{2}}\\&+\frac{\xi c^{-\xi\ln{2}}e^{1/\xi}}{(e^{1/\xi}-2)}\omega^{-1+\xi\ln{2}}|\ln{(\omega/c)|}\,.
\end{split}}
Since $\xi\ln2<1$ and we are interested in $\omega\ll 1$, the dominant term above is the last one such that we have
\eq{\Sigma_1\approx\frac{\xi c^{-\xi\ln{2}}}{1-2e^{-1/\xi}}\,\omega^{-1+\xi\ln{2}}|\ln{(\omega/c)|}\label{eq:S1-approx}}

Turning to the second summation in Eq. \ref{eq:pu-sum-two-terms},
\eq{\Sigma_2=\omega^{-\mu}\sum_{r=r_\ast(\omega)+1}^{L}N_r[\omega_\ast(r)]^{\mu-1}\,,}
it also splits into two terms $\Sigma_2 = \Sigma_2^{(1)}+\Sigma_2^{(2)}$ depending on the value of $N_r$,
\eq{
    \Sigma_2^{(1)}&=c^{\mu-1}\omega^{-\mu}\sum_{r_\ast(\omega)+1}^{L/2}(r-1)e^{-r[\ln{2}+(\mu-1)/\xi]}\,,\label{eq:s21}\\
    \Sigma^{(2)}_2 &=c^{\mu-1}\omega^{-\mu}\sum_{r=L/2+1}^{L}(L-r+1)e^{-r[\ln{2}+(\mu-1)/\xi]}\label{eq:s22}\,.
}
Note however that each summand in Eq.~\ref{eq:s22} is exponentially small in $L$ as $\xi\ln2>0$ and $\mu>1$ and but the number of terms in the sum is only linear in $L$. Hence $\Sigma_2^{(2)}$ is also exponentially small in $L$, and can be neglected such that $\Sigma_2\overset{L\to\infty}{\to}\Sigma_2^{(1)}$.
Each of the term in the latter sum is exponentially suppressed in system size(as, $\mu>1$). The sum $\Sigma^{(1)}_2$ in Eq.~\ref{eq:s21} can be performed exactly leading to 
\eq{\begin{split}
    \Sigma_2^{(1)}=& c^{\mu-1}\frac{e^{-\frac{L}{2\xi}(\mu-1+\xi\ln{2})}}{[2e^{(\mu-1)/\xi}-1]^2}[L e^{(\mu-1)/\xi}+L-1]\omega^{-c}\\&+\frac{c^{-\xi\ln{2}}}{[2e^{(\mu-1)/\xi}-1]^2}\omega^{-1+\xi\ln{2}}\\&+\frac{\xi c^{-\xi\ln{2}}}{[2e^{(\mu-1)/\xi}-1]}\omega^{-1+\xi\ln{2}}|\ln{(\omega/c)}|\,.
\end{split}}
The first term on the right-hand side of the equation above is again exponentially small in $L$ and hence can be neglected. Moreover, since we are interested in $\omega \ll 1$, the third term dominates over the second and we therefore have
\eq{
\Sigma_2 = \frac{\xi c^{-\xi\ln{2}}}{[2e^{(\mu-1)/\xi}-1]}\omega^{-1+\xi\ln{2}}|\ln{(\omega/c)}|\,.
\label{eq:S2-approx}
}
Using Eq.~\ref{eq:S1-approx} and Eq.~\ref{eq:S2-approx} we finally have
\eq{
    \langle\puo(\omega)\rangle\approx& C\,\omega^{-1+\xi\ln{2}}|\ln{(\omega/c)}|\,,
}
which is the result in Eq.~\ref{eq:pu-o-ans}. As a matter of completeness we note that the prefactor $C$ above is give by
\eq{
C=c^{-\xi\ln{2}}\xi\bigg[\frac{1}{1-2e^{-1/\xi}}+\frac{1}{2e^{(\mu-1)/\xi}-1}\bigg]\,,
}
which, reassuringly, is necessarily positive due to $\xi<1/\ln{2}$ and $\mu>1$.

\section{Logarithmic correction to $\langle{\pu(t)}\rangle$ \label{app:logcorr}}
At the end of Sec.~\ref{sec:log-growth}, we had attributed the discrepancy between the power-law exponent $a$ obtained from the fits to the numerical data in Fig.~\ref{fig:Pt-S2t} and those obtained from the theory, to logarithmic corrections in $t$ to $\expval{\pu(t)}$. Here we present the evidence for the same.

\begin{figure}[!h]
    \centering
    \includegraphics[width=\linewidth]{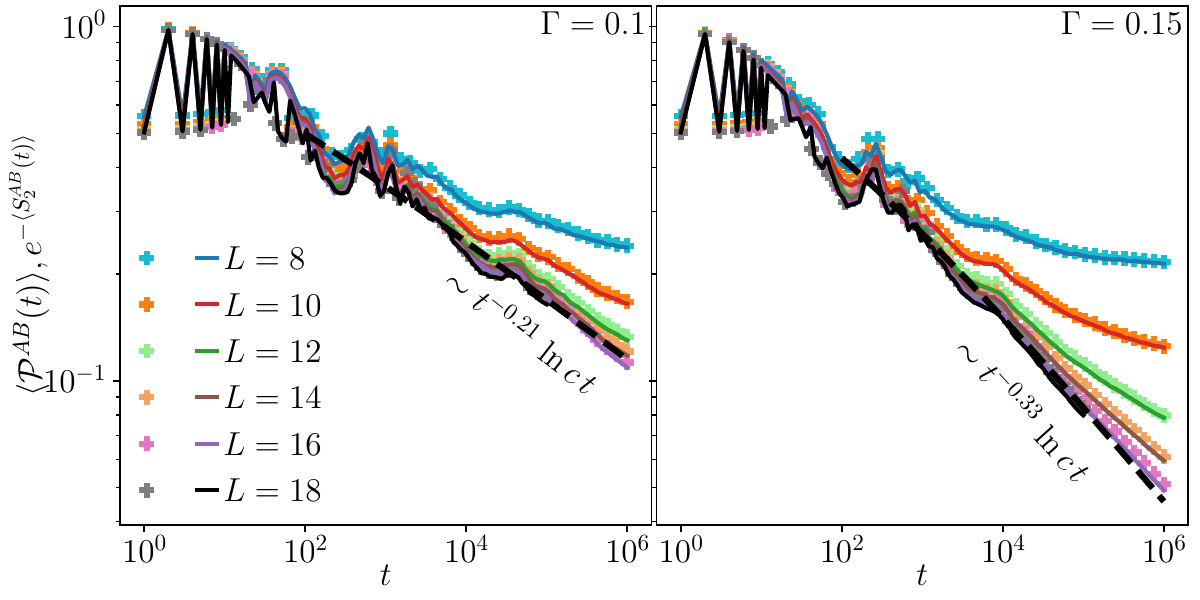}
    \caption{$\expval{\pu(t)}$ as a function of $t$ shows power-law exponents consistent with the theory when the logarithmic corrections are taken into account.}
    \label{fig:Pt-S2-log}
\end{figure}

According to our theoretical prediction in Eq.~\ref{eq:Pt-ans}, the dynamics of purity is given by
\eq{
\langle \pu(t) \rangle \sim t^{-\xi \ln 2} \ln c\, t = t^{-\xi \ln 2} (\ln c + \ln t)\,,
}
which contains the aforementioned logarithmic in $t$ correction.
We consider the raw data shown in Fig.~\ref{fig:Pt-S2t} and fit the above form. These fits are shown in Fig.~\ref{fig:Pt-S2-log}.
The power-law exponent so extracted is 
$a\approx 0.21$ and $a\approx 0.33$ for $\Gamma=0.1$ and 0.15 respectively. These are in excellent agreement with the theoreticak predictions (see Sec.~\ref{sec:log-growth} and therefore presents compelling evidence for the quantitative correctness of our theory as well.

\bibliography{refs}

\end{document}